\documentclass[twocolumn]{aastex631}

\newcommand\swinburne{Centre for Astrophysics and Supercomputing, Swinburne University of Technology, P.O. Box 218, Hawthorn, VIC 3122, Australia}

\newcommand\CSIRO{Australia Telescope National Facility, CSIRO, Space and Astronomy, PO Box 76, Epping, NSW 1710, Australia}
\newcommand{\OzGravMonash}{OzGrav: The Australian Research Council Centre of Excellence for Gravitational Wave Discovery, Clayton, VIC 3800, Australia}
\newcommand{\OzGravSwin}{OzGrav: The Australian Research Council Centre of Excellence for Gravitational Wave Discovery, Hawthorn, VIC 3122, Australia}
\newcommand\MQ{Department of Physics and Astronomy and MQ Research Centre in Astronomy, Astrophysics and Astrophotonics, Macquarie University, NSW 2109, Australia}
\newcommand\monash{School of Physics and Astronomy, Monash University, VIC 3800, Australia}

\begin{document}

\title{The gravitational-wave background null hypothesis: Characterizing noise in millisecond pulsar arrival times with the Parkes Pulsar Timing Array}

\author[0000-0002-2035-4688]{Daniel J. Reardon}\thanks{E-mail: dreardon@swin.edu.au}
\affiliation{\swinburne}
\affiliation{\OzGravSwin}

\author[0000-0002-9583-2947]{Andrew Zic}\thanks{E-mail: andrew.zic@csiro.au}
\affiliation{\CSIRO}
\affiliation{\MQ}

\author[0000-0002-7285-6348]{Ryan M. Shannon}
\affiliation{\swinburne}
\affiliation{\OzGravSwin}

\author[0000-0003-3432-0494]{Valentina Di Marco}
\affiliation{\monash}
\affiliation{\OzGravMonash}

\author[0000-0003-1502-100X]{George B. Hobbs}
\affiliation{\CSIRO}

\author[0009-0001-5071-0962]{Agastya Kapur}
\affiliation{\MQ}
\affiliation{\CSIRO}

\author[0000-0001-9208-0009]{Marcus E. Lower}
\affiliation{\CSIRO}

\author[0000-0001-5131-522X]{Rami Mandow}
\affiliation{\MQ}
\affiliation{\CSIRO}

\author[0000-0001-5532-3622]{Hannah Middleton}
\affiliation{School of Physics and Astronomy \& Institute for Gravitational
Wave Astronomy, University of Birmingham, Birmingham, B15 2TT, UK}

\author[0000-0002-5455-3474]{Matthew T. Miles}
\affiliation{\swinburne}
\affiliation{\OzGravSwin}

\author{Axl F. Rogers}
\affiliation{Institute for Radio Astronomy \& Space Research, Auckland University of Technology, Private Bag 92006, Auckland 1142, New Zealand}

\author[0009-0002-9845-5443]{Jacob Askew}
\affiliation{\swinburne}
\affiliation{\OzGravSwin}

\author[0000-0003-3294-3081]{Matthew Bailes}
\affiliation{\swinburne}
\affiliation{\OzGravSwin}

\author[0000-0002-8383-5059]{N. D. Ramesh Bhat}
\affiliation{International Centre for Radio Astronomy Research, Curtin University, Bentley, WA 6102, Australia}

\author[0000-0002-2037-4216]{Andrew Cameron}
\affiliation{\swinburne}
\affiliation{\OzGravSwin}

\author[0000-0002-0893-4073]{Matthew Kerr}
\affiliation{Space Science Division, US Naval Research Laboratory, 4555 Overlook Ave. SW, Washington DC 20375, USA}

\author[0000-0003-4847-4427]{Atharva Kulkarni}
\affiliation{\swinburne}
\affiliation{\OzGravSwin}

\author[0000-0001-9445-5732]{Richard N. Manchester}
\affiliation{\CSIRO}

\author[0000-0002-3922-2773]{Rowina S. Nathan}
\affiliation{\monash}
\affiliation{\OzGravMonash}

\author[0000-0002-1942-7296]{Christopher J. Russell}
\affiliation{CSIRO Scientific Computing, Australian Technology Park, Locked Bag 9013, Alexandria, NSW 1435, Australia}

\author[0000-0003-0289-0732]{Stefan Os{\l}owski}
\affiliation{Manly Astrophysics, 15/41-42 East Esplanade, Manly, NSW 2095, Australia}

\author[0000-0001-7049-6468]{Xing-Jiang Zhu}
\affiliation{Advanced Institute of Natural Sciences, Beijing Normal University, Zhuhai 519087, China}

\begin{abstract}

The noise in millisecond pulsar (MSP) timing data can include contributions from observing instruments, the interstellar medium, the solar wind, solar system ephemeris errors, and the pulsars themselves. The noise environment must be accurately characterized in order to form the null hypothesis from which signal models can be compared, including the signature induced by nanohertz-frequency gravitational waves (GWs).
Here we describe the noise models developed for each of the MSPs in the Parkes Pulsar Timing Array (PPTA) third data release, which have been used as the basis of a search for the isotropic stochastic GW background.
We model pulsar spin noise, dispersion measure variations, scattering variations, events in the pulsar magnetospheres, solar wind variability, and instrumental effects. We also search for new timing model parameters and detected Shapiro delays in PSR~J0614$-$3329 and PSR~J1902$-$5105. The noise and timing models are validated by testing the normalized and whitened timing residuals for Gaussianity and residual correlations with time. We demonstrate that the choice of noise models significantly affects the inferred properties of a common-spectrum process. Using our detailed models, the recovered common-spectrum noise in the PPTA is consistent with a power law with a spectral index of $\gamma=13/3$, the value predicted for a stochastic GW background from a population of supermassive black hole binaries driven solely by GW emission. 

\end{abstract}


\section{Introduction} \label{sec:intro}

Encoded in pulse arrival times are both deterministic and stochastic signals that each contain processes of fundamental interest.
A complete model of both is necessary for correct pulsar timing inference.
Deterministic terms include the motion of the Earth around the solar system barycenter, and, if the pulsar is in a binary system, the motion of the pulsar about its companion.  The study of binary systems and the accuracy of pulsar timing have enabled many tests of fundamental physical importance. These include stringent tests of relativistic gravity through the timing of double neutron star systems \cite[][]{Kramer+21} and constraints on the nuclear equations of state through the detection of Shapiro delays in pulsar--white dwarf binary systems \cite[][]{Demorest+10}.

Noise in pulsar timing observations can be described using physical or phenomenological models. 
Physical models are motivated by physical processes that can impact pulse arrival times, such as the interstellar medium or rotational instabilities in the neutron stars.
Phenomenological models are motivated by detection of processes in arrival time measurements and do not necessarily have a definitive physical origin. 
Processes are typically classified as white or red if they are uncorrelated or correlated between observations.  Processes can also be achromatic and equally affect received radiation at all observing radio frequencies $\nu$, or be strongly chromatic \cite[][]{Cordes+10}.

One of the most sought-after signals not yet detected in pulsar timing data sets is stochastic.  Nanohertz-frequency gravitational waves (GWs) passing through the galaxy alter the arrival times of pulses from pulsars in a spatially coherent manner. The most likely source of GWs is the incoherent stochastic superposition of GWs emitted by an ensemble of supermassive black hole binaries \citep[SMBHBs;][]{Rajagopal+95}. This GW background (GWB) is expected to manifest as a temporally correlated (red) stochastic processes in a pulsar timing data set.
In order to distinguish the GWB from other stochastic processes it is necessary to identify the unique Hellings--Downs (HD) correlations it imparts \citep[][]{HellingsDowns}.  
The GWB is expected to alter the arrival times of pulsars by only tens to hundreds of nanoseconds. For both of these reasons it is necessary to monitor an ensemble of millisecond pulsars (MSPs), referred to as a pulsar timing array \citep[PTA;][]{Foster+90a}, as these pulsars can be timed to the highest precision and are the most inherently rotationally stable. MSPs are believed to be formed in close binary systems by a process known as ``recycling" \citep{Bhattacharya+91}.

Decades-long pulsar timing experiments have been ongoing in Australia \citep[tthe Parkes PTA (PPTA);][]{Manchester+13}, Europe \citep[the European PTA (EPTA);][]{Kramer+13}, and in North America \citep[North American Nanohertz Observatory for Gravitational Waves (NANOGrav);][]{Demorest+13}. These groups, in addition to more recent projects established in China \citep[Chinese PTA (CPTA);][]{KJLee16} and India \citep[Indian PTA (InPTA);][]{inpta_dr1}, and with the MeerKAT radio telescope in South Africa \citep[MeerKAT PTA (MPTA)][]{mpta} and the Fermi-Large Area Telescope \citep{fermi+22}, form the set of global experiments searching for nanohertz-frequency GWs. The International Pulsar Timing Array \citep[IPTA;][]{Hobbs+10}, comprising several of these experiments (EPTA, InPTA, NANOGrav, PPTA), improves the sensitivity to GWs by combining its constituent data sets.

In order to detect the background, it is imperative to fully characterize the pulsar timing data sets. This includes modeling the myriad of noise sources, many of which are astrophysical foregrounds that must be characterized.

The emitting neutron star itself contributes both white and red noise to the pulse arrival times.
Individual pulses vary in intensity and morphology from pulse to pulse, causing pulse shape variations known as jitter (some relativistic systems also show profile variations due to relativistic precession effects). Most of the variations appear to be independent from pulse to pulse, which contributes to excess white noise in pulse time-of-arrival (TOA) measurements \cite[][]{Oslowski+11,Shannon+14}. There are many examples of nonrecycled pulsars that show pulse shape variations on long time scales.  There are a few examples of this amongst the MSPs \cite[][]{Shannon+16, Jennings+22}. 
Spin noise, i.e., instabilities in the apparent rotation rate of neutron stars, is the dominant stochastic process in nonrecycled pulsars.  The presence of spin noise has been reported across the MSP population \citep{IPTA_dr1_noise, PPTA_dr2_noise}.  The spectral shape and the amplitude of the red noise may be comparable to that expected from the GWB \cite[][]{Shannon+10}. 

The interstellar medium can also introduce stochastic variations to pulse arrival times. The ionized component of the interstellar medium (IISM) is thought to be highly turbulent, with the largest-scale variations driven by supernova explosions and winds in star-forming clusters, and with a turbulent cascade producing density fluctuations with structures as small as $\sim 1$\,au \citep{Armstrong+77}. As the column density of the plasma varies between the pulsar and the Earth (because of the transverse motion of the line of sight), a number of time-varying effects are potentially measurable in the pulse arrival times. Variations in the total electron column density \cite[dispersion measure (DM);][]{Keith+13} induce a signal delay $\propto \nu^{-2}$. The inhomogeneities of the turbulence result in multipath propagation and diffractive and refractive scattering of the pulsar radiation. These can both distort the pulse shape and cause arrival time variation \cite[][]{Cordes+16, Shannon+17}.

In addition to the GWB, there are other credible sources for stochastic processes that are correlated between pulsars and likely to be present in pulsar timing data sets at some level. An error in the time referencing will result in arrival time variations that are strongly correlated between pulsars \cite[][]{Hobbs+10}.  Errors in the barycentering of arrival times (due to an incorrect model of the solar system) will manifest as dipolar-correlated arrival time variations \cite[][]{Champion+10, Bayesephem}.  Both of these noise sources could cause temporal correlations with similar amplitude and spectral shape to that of the GWB \cite[][]{Tiburzi+16, NG11yrSGWB, Bayesephem}. Unmodeled variations in the solar wind also manifest themselves as excess DM. These could also impart broadly dipolar spatial correlations. The temporal correlation of this signal is expected to be different from that of the GWB  \citep{Tiburzi+16}. 

Instrumentation potentially can also introduce excess noise \citep{IPTA_dr1_noise}.  Over the course of PTA experiments, instrumentation is often upgraded.  Delays introduced in the instrumentation may not be known a priori, and if offsets are incorrectly applied between different instruments, the small step changes may appear as a red-noise-like process. Changes in the polarization response of a telescope receiving system can also distort pulse profiles and result in temporally correlated noise \cite[][]{vanStraten06, vanStraten13}. 

Accurately characterizing the noise is crucial for detecting a GWB. Noise mis-specification could result in insensitive GW searches or the nondetection of a GWB when one was present. It could also potentially result in the false detection of a background in data containing noise. Noise analyses of individual pulsars have been conducted separately from (and jointly with) searches for GWs by the EPTA \citep{EPTA_dr2_noise}, the InPTA \citep{InPTA_noise}, NANOGrav \citep{NG12.5yrSGWB}, the PPTA \citep{PPTA_dr2_noise}, and their union in the IPTA \citep{IPTA_dr1_noise}. 

In this paper, we present noise analyses for the MSPs in the PPTA third data release (PPTA-DR3). 
This work is part of a set of PPTA papers, which includes a description of the data release in \citet{PPTA-DR3_data}, and a search for the isotropic stochastic GWB in \citet{PPTA-DR3_gwb}. We describe our methodology for identifying and characterizing noise sources in Section \ref{sec:models} and present and interpret the preferred models in Section \ref{sec:results}. We summarize the impact of these noise models in Section \ref{sec:discussion} and draw our conclusions in Section \ref{sec:conclusions}.

\section{Methods and noise model components} \label{sec:models}

The data set used for this analysis is described in \citet{PPTA-DR3_data}, including the pulsar ephemerides and TOAs that form the basis for the noise modeling described in this work. We fit initial timing models using \textsc{Tempo2} \citep{Edwards+06}, beginning from the timing analyses of the previous data releases \citep{Reardon+16, Reardon+21}. For new pulsars added to the PPTA since the second data release (PPTA-DR2), we use the initial timing models from \citet{Curylo+23}. For four pulsars we required updates to the timing models. However, the timing model parameters are treated as nuisance parameters for GW searches, and as such they are analytically marginalized in this work. In this section we describe the construction of our noise model of deterministic and stochastic processes that are not accounted for in the timing model. 

Bayesian inference is used to measure the noise parameters, as detailed in our companion GWB analysis paper \citep[][and references therein]{PPTA-DR3_gwb}. In brief, the timing residuals are modeled with a Gaussian likelihood \citep{vanHaasteren+09}. Time-correlated (red) stochastic processes are modeled in the time domain as Gaussian processes \citep{Lentati+13, vanHaasteren+14} using Fourier basis functions. The Fourier amplitudes can be constrained to follow a distribution such as a power law, where the amplitude and spectral index are free parameters, but the Fourier amplitudes are analytically marginalized along with the timing model. The posterior probabilities of the noise model parameters are evaluated from Bayes's theorem using the \textsc{enterprise} package \citep{Enterprise} and a Markov Chain Monte Carlo algorithm with parallel tempering \citep[\textsc{ptmcmcsampler};][]{PTMCMC}. In this framework, model comparison can be achieved through the Bayes factor ($\mathcal{B}$), but for this work we are not as concerned with the model support from the data as we are with including noise terms to reduce the risk of model mis-specification during a search for common processes, including the GWB. For example, it is common practice \citep{NG12.5yrSGWB, PPTA_dr2_crn, EPTA_dr2_crn, IPTA_dr2_gwb} to include models for both red achromatic noise and DM variations in all pulsars, regardless of the evidence for such processes from the data themselves. The reason is that these processes, as well as others, must be present in the data based on physical arguments, although the level of their contribution is not known a priori.

PTA data are highly complex, and the noise processes present within the data are not fully understood. For our work, we take a liberal approach with the addition of noise terms to describe potential processes in the residuals. Our motivation is primarily to mitigate issues arising from unmodeled noise terms that may ``leak'' into the signals of interest. This can lead to inaccurate characterization or, at worst, false detections of such signals. On the other hand, if the presence of a noise term is not statistically supported by the data, then the parameters describing that process tend to be unconstrained, possibly below some upper bound, and display little to no covariance with other parameters. The result is that the inclusion of these models has little impact on the parameter estimation of the signals of interest. The most conservative approach in a Bayesian framework \citep[except in an upper-limit regime;][]{Hazboun+20} would be to include models describing all conceivable noise processes in the data, allowing the data to select the levels of the noise terms that describe it best. We cannot take this approach because it is too computationally expensive at present. Furthermore, there are likely noise processes present in the data that have not yet been identified and described. Instead, we consider noise terms that have been found in historical analyses of the PPTA \citep{PPTA_dr2_noise} and IPTA \citep{IPTA_dr1_noise} analyses, while also searching for new terms using the latest ultra--wide-bandwidth data.

\subsection{White noise}

Pulsar TOAs are measured by cross-correlating the observed pulse profile with a standard template. If the recorded profiles contain only radiometer noise and the template is accurate, then the uncertainties associated with the TOAs are accurate. However, when there are other factors present such as residual radio frequency interference (RFI), changes in the pulse profile with time, instrumental artifacts, or template errors, then the uncertainty estimations will not be correct. To account for these issues, white (uncorrelated) noise parameters are required.

Three white-noise parameters are used to describe excess uncorrelated noise in the PPTA-DR3: EFAC ($F$), EQUAD ($Q$), and ECORR ($E_c$), as defined in other pulsar timing noise analyses \citep[e.g.][]{temponest, NG12.5yrSGWB}. $F$ is a scale factor to the TOA uncertainties, and $Q$ is an excess noise added in quadrature. The modified uncertainties are $\sigma_t = ((F\sigma_{t,0})^2 + Q^2)^{1/2}$, for original uncertainty $\sigma_{t,0}$. $E_c$ is an additional noise term added in quadrature that describes noise that is completely correlated between simultaneous observations at different frequencies, while being completely uncorrelated between epochs \citep[described in Appendix C of][]{ng9yr_data}. $E_c$ accounts for the fact that the subbanded TOAs in frequency are not independent, primarily because of pulse jitter \citep{Oslowski+11, Shannon+14, Lam+19, Parthasarathy+21}.

The jitter noise modeled by $E_c$ is expected to decorrelate over a wide bandwidth \citep[e.g. Figure 3 of][]{Parthasarathy+21}. In the PPTA-DR2, one $E_c$ parameter was used for each of three observing bands, but it was assumed that each of the bands was sufficiently independent. Data from the UWL receiver include each of these bands as a subset of a continuous band from 704 to 4032\,MHz (see \citet{PPTA-DR3_data} for more details). A description of an $E_c$ parameter that accounts for the decorrelation as a function of frequency is deferred to future work (A.~Kulkarni et al. 2023, in preparation). For our analysis, we approximate this decorrelation by using three $E_c$ parameters across the UWL band, centered near the discrete bands from the PPTA-DR2 (i.e. $\nu<960\,$MHz, $960\,$MHz$ < \nu < 2048\,$MHz, and $\nu > 2048\,$MHz). We additionally include a global $E_c$ parameter for the whole UWL band, which models any broadband jitter noise, or low-level instrumental offsets.

\subsection{Timing noise and dispersion measure variations}

Temporal variations in DM and spin noise require careful characterization when searching for correlated signals across a PTA. In this analysis, we include a power-law model to describe the timing noise and DM variations for every pulsar. As described above, this choice is physically motivated: variations in the turbulent interstellar medium and pulsar spin irregularities do occur and will therefore influence the timing residuals even if at marginal levels. 

The number of Fourier frequencies used in the bases employed to model DM variations and spin noise are determined by the time span for each pulsar and the highest fluctuation frequency we model. For the achromatic red process, we model up to a maximum frequency of $1/(240\,\mathrm{days})$, while for DM variations we model up to $1/(60\,\mathrm{days})$. These maximum frequencies were chosen following an initial analysis with a broken power-law model, which determines the frequency above which the spectral index flattens \citep[see ][]{NG12.5yrSGWB}. We found that most pulsars become insensitive to such a break once the power law approaches the white-noise level. We chose these values for the maximum frequency based on when this condition occurs for all pulsars.

We model the achromatic red noise with a power-law power spectral density (PSD):
\begin{equation}
    \label{eq:rn_spec}
    P_\mathrm{Red}(f; A, \gamma ) = \frac{A^2}{12\pi^2} \left( \frac{f}{f_\mathrm{yr}}\right)^{- \gamma} ~(\mathrm{yr}^{3}),
\end{equation}
where $A$ is the amplitude, $\gamma$ is the spectral index defined such that red processes have a positive index, and $f$ is the fluctuation frequency. The amplitude $A$ is a dimensionless strain derived from a GW amplitude spectrum of the form $h_c = A(f/1 {\rm yr}^{-1})^{\alpha}$, where $\alpha = (3 - \gamma)/2$.  DM variations are similarly modeled, but the amplitude of the PSD is scaled as $P_\mathrm{DM}(f, \nu; A^{\rm DM}, \gamma^{\rm DM}) = (\nu/1400\,\mathrm{MHz})^{-2}P_\mathrm{Red}(f; A^{\rm DM}, \gamma^{\rm DM} )$. We set the priors for the parameters of each Gaussian process with a PSD derived from Equation \ref{eq:rn_spec} to uniform distributions ($\mathcal{U}$) in the ranges $\pi(\log_{10}A) = \mathcal{U}[-18, -11]$ and $\pi(\gamma) = \mathcal{U}[0, 7]$.
    
 We also searched for an additional, high fluctuation frequency (HFF) achromatic red-noise process for some pulsars, modeling Fourier frequencies up to $1/(30\,\mathrm{days})$. Our primary motivation was to capture shallow-spectrum achromatic red-noise processes. Steep-spectrum red-noise processes dominate the lowest Fourier harmonics, and so models that only consider low harmonics will be dominated by any steep-spectrum process(es). In the presence of a steep common-spectrum process such as a GWB, precisely timed pulsars may exhibit noise originating from processes other than the common process at high fluctuation frequencies. The single-pulsar noise analyses do not assume a common process (which requires the PTA as a whole), and so two red-noise processes may be required to adequately describe the noise present in the data. We ultimately include this HFF red-noise term in the model for pulsars if the spectral properties are constrained and not completely degenerate with the nominal red-noise process.

\subsection{Scattering, band, and system noise}

For pulsars with a large DM or high timing precision, we search for scattering noise, which scales with radio frequency approximately as $\nu^{-4}$ \citep{Lang71}. The PSD for this scattering noise, modeled as a Gaussian process, is therefore $P_\mathrm{Chr}(f, \nu; A^\mathrm{Chr}, \gamma^\mathrm{Chr}) = (\nu/1400\,\mathrm{MHz})^{-4}P_\mathrm{Red}(f; A^\mathrm{Chr}, \gamma^\mathrm{Chr} )$. While $\nu^{-4}$ is an appropriate model for the frequency scaling at the required precision, individual sources have been observed to scale differently \citep{Geyer+16}, which may manifest as excess noise, particularly at low radio frequencies. 

Excess noise in isolated observing systems and frequency bands has been observed for many pulsars but is poorly understood \citep[e.g.][]{IPTA_dr1_noise, PPTA_dr2_noise}. The origins of such noise could include residual RFI, secondary effects from interference flagging (e.g., flagging leading to subtle changes in the effective observing frequency), unmodeled system offsets, pulse profile variability, calibration errors \citep{vanStraten13}, scintillation interacting with template errors, or errors in the correction of DM and scattering variations \citep{Cordes+16}, and possibly the interaction between any of these effects.

We allow all pulsars with modeled scattering noise to have band noise at low frequencies (modeled as a Gaussian process with PSD of Equation \ref{eq:rn_spec}), which accounts for any excess noise induced, for example, by errors in the assumed IISM noise frequency scaling. We additionally allow for low-frequency band noise in the highest-precision pulsars, which are most sensitive to the various potential sources of such noise. We also include ``mid''-frequency ($960\,\text{MHz} < \nu \leq 2048\,\text{MHz}$) and ``high''-frequency ($\nu > 2048$\,MHz) band-noise terms for PSR~J0437$-$4715, to capture higher-order frequency-dependent noise. This frequency-dependent noise is most apparent in PSR~J0437$-$4715 \citep{PPTA_dr2_noise} because of its brightness, which makes it more sensitive to profile stochasticity, instrumental effects not well described by a single system noise term, and IISM effects. We model band noise as a power-law red-noise process, with PSD described by Eq. \ref{eq:rn_spec}, but operating on TOAs selected by frequency according to the specifications above.

We searched for system noise in each pulsar by performing parameter estimation for a power-law red-noise process, with power spectral density described by Eq. \ref{eq:rn_spec}. These red-noise processes operated only on subsets of TOAs selected by the \texttt{-group} flag on the data, which specifies the receiver and signal processing system used for a given TOA measurement. We examined the marginal posteriors for each system noise log-amplitude and only retained system noise terms where there was an increased posterior density over the density in the low-amplitude posterior tail (corresponding to estimated Savage--Dickey Bayes factors $\log\mathcal{B} \gtrsim 1$). After this selection process, we found that several system noise terms had maximum likelihood spectral indices consistent with $\gamma = 0$, suggesting that these systems may have an associated time-uncorrelated noise component. To account for this, we included an additional $E_c$ parameter for these systems and did not model their system noise as a red-noise process.

\subsection{Instrumental timing offsets}

Accurate correction for timing offsets is crucial for any inference from pulsar timing. Because of their origin in the telescope signal chain, they usually affect many (or all) pulsars in a PTA. A sequence of irregularly spaced timing offsets of varying magnitude can mimic a power-law process and will induce monopole-correlated signal across the PTA. Left unmitigated, these offsets may dominate interpulsar correlated signals of astrophysical origin. On the other hand, indiscriminate identification and correction of timing offsets may falsely whiten the timing residuals for a pulsar, removing any astrophysical signal within.

While timing offsets within the PPTA-DR2 have been scrutinized \citep{ppta_dr2}, it is important to characterize potential offsets in the UWL/Medusa system. To ensure a more complete accounting of all timing offsets, we implemented a timing offset search method in our noise modeling procedure. This was implemented via parameter estimation for a time-domain waveform described by a heaviside unit step function $H$ in single-pulsar noise modeling 
\begin{equation}
    J = \mathrm{sgn}(s) {A_\mathrm{JUMP}} H(t - t_\mathrm{JUMP})\mathrm{,}
\end{equation}
where $s$ is a free parameter ranging between $\pm 1$ that describes the sign of the timing offset, $A_\mathrm{JUMP}$ is the timing offset amplitude, and $t_\mathrm{JUMP}$ is the epoch of the timing offset. The prior ranges used for these parameters were $s \in \mathcal{U}[-1,1]$, $\log_{{10}} A_\mathrm{JUMP} \in \mathcal{U}[-10,-6]$, and $t_\mathrm{JUMP} \in \mathcal{U}[t_i, t_j]$, where $t_i$, $t_j$ are the boundaries of successive, overlapping 243-day windows that cover the data set.

We first performed the timing offset search and parameter estimation on individual pulsars. To ensure complete coverage and to avoid the parameter estimation being dominated by a small number of significant timing offsets, we searched for individual timing offsets in discrete 243-day time windows, which overlapped by 30 days with their adjacent windows. This was chosen so that each window spanned 183 days (approximately 6 months), while ensuring reliable detection of any timing offsets close to the 6-month boundaries.

To determine whether any apparent timing offsets were common among the pulsars, we inspected and combined the individual pulsar timing offset posteriors using a factorized likelihood approach \citep{Taylor+22}. We only considered and corrected for timing offsets that had statistical support from a plurality of pulsars. We validated this approach by removing the JUMP parameter for a timing offset occurring on MJD 59200 from the timing model parameter files and searching for that offset with this approach. We detected the relevant timing offset confidently.

Future improvements to this approach could include applying the method to subsets of the data (e.g., applying the method to TOAs from instrumental subbands), or implementation as a monopole-correlated common signal in full PTA analysis, which may improve sensitivity. We also note that astrophysical signals such as GW bursts with memory \citep{cordesbwm,ngbwm,pptabwm} may mimic a timing offset. We do not search for such signals in our work, so revisiting the timing offset measurements may be necessary for future searches for GW memory events.

\subsection{Magnetospheric, interstellar, and Other deterministic events.}

While it is often assumed that DM variations follow a Gaussian power-law process, this need not be the case. Plasma intermittency could cause departures from such a process, as could the presence of discrete structures in the IISM. Consequently, in addition to the Gaussian process DM variations, we include a Gaussian-shaped DM event for PSR~J1603$-$7202 \citep[Equation 7 of][]{PPTA_dr2_noise}, to describe its extreme scattering event \citep{Coles+15, Reardon+23} and annual DM variations \citep[Equation 8 of][]{PPTA_dr2_noise} for PSR~J0613$-$0200 \citep{Keith+13}.

Four pulsars show evidence for events in their magnetosphere, characterized by a sudden frequency-dependent offset in the timing residuals, with an exponential-like decay due to time- and frequency-dependent pulse shape changes. PSR~J1713+0747 showed two such events across our data set \citep{Demorest+13,Lam+18}, as its recent third event \citep{Singha+21,Jennings+22} was excluded from our analysis. These observed pulse profile shape change events are modeled as a chromatic step function with an exponential recovery \citep[Equation 6 of][]{PPTA_dr2_noise}. We include one for the strong event in PSR~J1643$-$1224 \citep{Shannon+16}, two for PSR~J1713$+$0747, and one each for PSR~J0437$-$4715 and PSR~J2145$-$0750 \citep{PPTA_dr2_noise}. Additionally, during the course of our analyses, we identified a Gaussian-like feature in the residuals of PSR~J1600$-$3053, spanning $\sim$months. This feature is not well described by other noise processes in the model and is unlikely to be related to the IISM because it is only apparent in the 20\,cm band (approximately $1$--$2$\,GHz). We modeled this feature with a time-domain Gaussian waveform of time delays ($t_{\rm Gauss}$) subtracted from the TOAs in this band:
\begin{equation}
t_{\rm Gauss}(t) = A_g \mathrm{exp}\left(\frac{(t - t_{g,0})^2}{2\sigma_g^2} \right)
\end{equation}
where $A_g$ is the amplitude of the feature in seconds, $t_{g,0}$ is the central epoch (MJD), and $\sigma_g$ is the width in days. We measure $\log_{10} (A_g / \mathrm{s})  = -5.83^{+0.10}_{-0.16}$, $t_{g,0} = 57575^{+8}_{-7}$, and $\log_{10} (\sigma_g / \mathrm{d})= 1.46\pm0.14$. An astrophysical origin (e.g. profile shape change) could be confirmed with detection in other PTAs.

\subsection{Noise model validation}

To assess the completeness of our noise models, we require that the noise-subtracted (whitened) and the band-averaged normalized residuals are consistent with white noise with zero mean and unit variance. We computed the Anderson--Darling statistic \citep[ADS;][]{Anderson+54} to test consistency between these whitened, normalized residuals and the expected standard normal distribution \citep[following previous PPTA noise analyses;][]{Reardon+16, PPTA_dr2_noise}.

For each observing band of each pulsar, we also conduct a least-squares spectral analysis of the whitened and normalized residuals \citep[forming the Lomb--Scargle periodogram;][]{Lomb76, Scargle82} and test whether the power at the lowest fluctuation frequencies ($f_t < 1/240\,{\rm days}$) is consistent with white noise. We test these frequencies because they are used for inference of the GWB signal in the companion analysis \citep{PPTA-DR3_gwb}.

\section{Results}
\label{sec:results}

\begin{figure*}
\centerline{\includegraphics[width=0.84\textwidth, trim= 0 0 0 0 , clip]{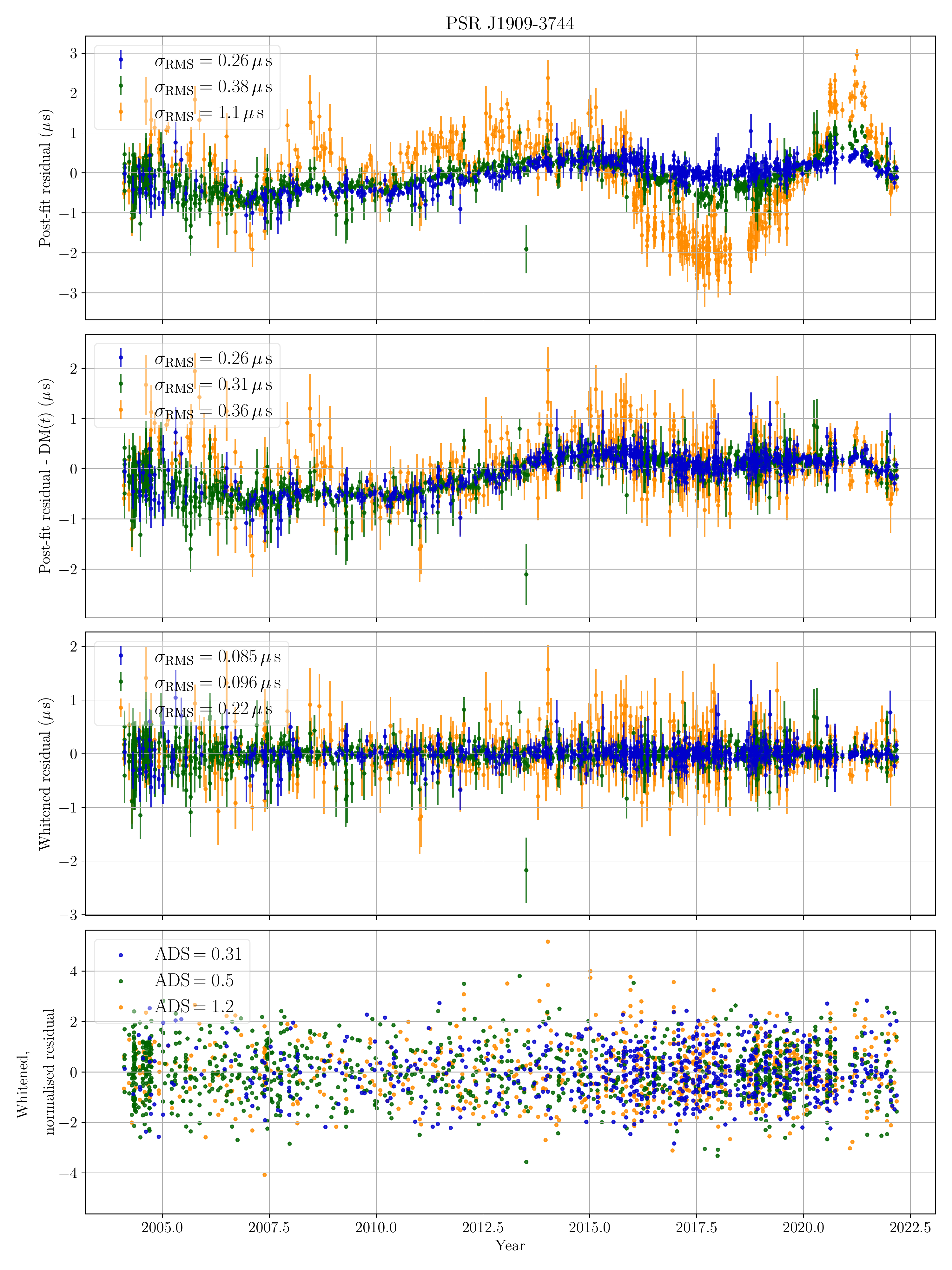}}
\caption{Band-averaged timing residuals and noise model validation for PSR~J1909$-$3744. Residuals from observations with radio wavelengths near 10, 20, and 40\,cm are shown in blue, green, and orange, respectively. The top panel shows the timing residuals with all noise processes present. The second panel shows the residuals after subtracting the maximum likelihood realization of the DM Gaussian process. The third and fourth panels show the residuals with the maximum likelihood realizations of all noise processes subtracted. Additionally, the residuals in the fourth panel have been normalized by the uncertainties. The legends of the top three panels show the weighted rms residual for each band, while the bottom panel shows the ADS.}
\label{fig:J1909_res}
\end{figure*}

The measured parameters for the most prevalent noise processes in our data set are shown in Table \ref{tab:noise_params}. We excluded PSR~J1741$+$1351 from this analysis and the subsequent GWB search \citep{PPTA-DR3_gwb} because we only have 16 unique observations of this pulsar in the data set, which is insufficient for modeling noise processes. The timing residuals for PSR~J1909$-$3744, with and without noise processes subtracted, are shown in Figure \ref{fig:J1909_res}. This pulsar is the most sensitive in the PPTA because of its low rms timing residuals and low level of jitter noise.

\begin{table*}
\centering
\caption{Measured parameters for processes in the PPTA-DR3 included in the noise models for multiple pulsars. Note: The parameter values are the medians, with uncertainties showing the central 68\% credible interval. The parameter names refer to the power spectrum index ($\gamma$) and amplitude ($A$) for achromatic red noise (Red), dispersion measure (DM) variations, high fluctuation frequency (HFF), Chromatic (Chr), and low-frequency ($\nu \leq 960\,$MHz) band noise (BN), along with the mean solar wind density at 1\,au ($n_e^\mathrm{SW}$).}
{\footnotesize
\begin{tabular}{lrrrrrrrrrrr}
\hline
\hline
\noalign{{\medskip}} 
    PSR Name  & $\gamma^{{\rm Red}}$& $\log_{{10}} A^{{\rm Red}}$& $\gamma^{{\rm DM}}$& $\log_{{10}} A^{{\rm DM}}$& $\gamma^{{
\rm HFF}}$& $\log_{{10}} A^{{\rm HFF}}$& $\gamma^{{\rm Chr}}$& $\log_{{10}} A^{{\rm Chr}}$& $\gamma^{{
\rm BN}}_{{\rm low}}$& $\log_{{10}} A^{{\rm BN}}_{{\rm low}}$& $n^{{\rm sw}}_e$ \\ 
\noalign{{\smallskip}}  \hline \noalign{{\medskip}} 
J0030+0451 & $3.4^{+2.4}_{-2.3}$  & $-16.3^{+2.5}_{-2.5}$  & $3.4^{+2.4}_{-2.3}$  & $-16.8^{+2.2}_{-2.2}$  & --  & --  & --  & --  & --  & --  & $6.0^{+1.4}_{-1.4}$  \\ \noalign{{\smallskip}} 
J0125$-$2327 & $3.3^{+2.5}_{-2.3}$  & $-16.9^{+2.1}_{-2.1}$  & $3.2^{+1.5}_{-1.0}$  & $-13.4^{+0.1}_{-0.2}$  & --  & --  & --  & --  & --  & --  & $2.3^{+2.8}_{-1.6}$  \\ \noalign{{\smallskip}} 
J0437$-$4715 & $3.3^{+0.9}_{-0.6}$  & $-14.3^{+0.2}_{-0.3}$  & $2.5^{+0.1}_{-0.1}$  & $-13.48^{+0.04}_{-0.04}$  & $0.5^{+0.4}_{-0.3}$  & $-14.3^{+0.1}_{-0.1}$  & $3.0^{+0.5}_{-0.4}$  & $-14.4^{+0.1}_{-0.1}$  & $2.9^{+2.4}_{-1.7}$  & $-16.6^{+2.0}_{-2.3}$  & $3.7^{+4.3}_{-2.6}$  \\ \noalign{{\smallskip}} 
J0613$-$0200 & $5.9^{+0.8}_{-1.1}$  & $-15.4^{+0.6}_{-0.5}$  & $2.4^{+0.3}_{-0.3}$  & $-13.6^{+0.1}_{-0.1}$  & --  & --  & $5.2^{+1.3}_{-1.9}$  & $-15.8^{+1.1}_{-0.8}$  & $2.8^{+2.5}_{-2.0}$  & $-17.4^{+2.1}_{-1.8}$  & $1.3^{+1.3}_{-0.9}$  \\ \noalign{{\smallskip}} 
J0614$-$3329 & $3.2^{+2.5}_{-2.3}$  & $-16.4^{+2.4}_{-2.4}$  & $5.0^{+1.4}_{-2.0}$  & $-13.8^{+0.5}_{-0.5}$  & --  & --  & --  & --  & --  & --  & $11.2^{+6.1}_{-7.1}$  \\ \noalign{{\smallskip}} 
J0711$-$6830 & $1.2^{+0.7}_{-0.6}$  & $-13.1^{+0.1}_{-0.2}$  & $3.2^{+2.4}_{-1.1}$  & $-14.1^{+0.6}_{-1.8}$  & --  & --  & --  & --  & --  & --  & $9.8^{+6.8}_{-6.7}$  \\ \noalign{{\smallskip}} 
J0900$-$3144 & $3.4^{+2.4}_{-2.3}$  & $-16.5^{+2.4}_{-2.4}$  & $2.0^{+1.8}_{-0.8}$  & $-12.7^{+0.2}_{-2.6}$  & --  & --  & --  & --  & --  & --  & $8.3^{+7.4}_{-5.9}$  \\ \noalign{{\smallskip}} 
J1017$-$7156 & $3.2^{+2.5}_{-2.1}$  & $-16.1^{+2.1}_{-2.6}$  & $2.3^{+0.2}_{-0.2}$  & $-12.89^{+0.04}_{-0.04}$  & $0.9^{+0.4}_{-0.5}$  & $-13.4^{+0.1}_{-0.2}$  & $0.8^{+0.6}_{-0.5}$  & $-13.7^{+0.1}_{-0.1}$  & $3.0^{+2.5}_{-2.1}$  & $-17.2^{+2.0}_{-1.9}$  & $10.4^{+6.5}_{-6.9}$  \\ \noalign{{\smallskip}} 
J1022+1001 & $3.2^{+2.4}_{-2.1}$  & $-16.6^{+2.1}_{-2.3}$  & $2.5^{+1.0}_{-0.7}$  & $-13.8^{+0.3}_{-0.4}$  & $1.9^{+3.1}_{-1.3}$  & $-15.0^{+1.9}_{-3.3}$  & --  & --  & --  & --  & $9.0^{+0.6}_{-0.6}$  \\ \noalign{{\smallskip}} 
J1024$-$0719 & $3.1^{+2.5}_{-2.1}$  & $-17.1^{+2.1}_{-2.0}$  & $3.5^{+1.0}_{-0.7}$  & $-13.9^{+0.3}_{-0.5}$  & --  & --  & --  & --  & --  & --  & $4.3^{+2.4}_{-2.3}$  \\ \noalign{{\smallskip}} 
J1045$-$4509 & $1.4^{+3.2}_{-1.0}$  & $-14.5^{+1.9}_{-3.8}$  & $2.9^{+0.2}_{-0.2}$  & $-12.38^{+0.04}_{-0.04}$  & --  & --  & $3.4^{+2.0}_{-1.4}$  & $-14.2^{+0.7}_{-1.2}$  & $2.9^{+2.5}_{-2.0}$  & $-17.0^{+2.3}_{-2.0}$  & $7.4^{+7.4}_{-5.3}$  \\ \noalign{{\smallskip}} 
J1125$-$6014 & $3.9^{+1.6}_{-1.4}$  & $-14.2^{+0.5}_{-0.7}$  & $3.6^{+0.3}_{-0.3}$  & $-13.2^{+0.1}_{-0.1}$  & --  & --  & --  & --  & --  & --  & $12.1^{+5.2}_{-6.5}$  \\ \noalign{{\smallskip}} 
J1446$-$4701 & $3.1^{+2.5}_{-2.1}$  & $-17.0^{+2.1}_{-2.0}$  & $2.7^{+2.7}_{-1.9}$  & $-16.9^{+2.5}_{-2.1}$  & --  & --  & --  & --  & --  & --  & $6.5^{+5.3}_{-4.2}$  \\ \noalign{{\smallskip}} 
J1545$-$4550 & $3.3^{+2.4}_{-2.2}$  & $-16.6^{+2.0}_{-2.3}$  & $4.4^{+0.9}_{-0.8}$  & $-13.7^{+0.2}_{-0.3}$  & --  & --  & --  & --  & --  & --  & $2.4^{+2.7}_{-1.7}$  \\ \noalign{{\smallskip}} 
J1600$-$3053 & $3.4^{+2.1}_{-2.1}$  & $-15.9^{+1.6}_{-2.8}$  & $2.3^{+0.3}_{-0.2}$  & $-13.2^{+0.1}_{-0.1}$  & $2.1^{+2.9}_{-1.4}$  & $-14.4^{+0.9}_{-3.3}$  & $1.5^{+0.7}_{-0.5}$  & $-13.8^{+0.1}_{-0.4}$  & $2.4^{+2.7}_{-1.4}$  & $-16.7^{+3.4}_{-2.3}$  & $3.4^{+0.8}_{-0.8}$  \\ \noalign{{\smallskip}} 
J1603$-$7202 & $2.9^{+2.5}_{-2.0}$  & $-17.2^{+2.1}_{-1.9}$  & $2.3^{+0.3}_{-0.2}$  & $-13.2^{+0.1}_{-0.1}$  & --  & --  & --  & --  & --  & --  & $4.2^{+4.9}_{-3.0}$  \\ \noalign{{\smallskip}} 
J1643$-$1224 & $0.6^{+0.4}_{-0.4}$  & $-12.7^{+0.1}_{-0.1}$  & $2.3^{+0.3}_{-0.2}$  & $-12.9^{+0.1}_{-0.1}$  & --  & --  & $0.8^{+0.4}_{-0.4}$  & $-13.2^{+0.1}_{-0.1}$  & $2.1^{+0.3}_{-0.3}$  & $-12.3^{+0.1}_{-0.1}$  & $4.5^{+1.5}_{-1.5}$  \\ \noalign{{\smallskip}} 
J1713+0747 & $2.9^{+2.5}_{-2.0}$  & $-17.5^{+1.9}_{-1.7}$  & $2.1^{+0.4}_{-0.3}$  & $-13.9^{+0.1}_{-0.1}$  & $0.7^{+3.3}_{-0.5}$  & $-14.5^{+0.3}_{-3.8}$  & --  & --  & $3.1^{+1.0}_{-0.8}$  & $-13.8^{+0.3}_{-0.4}$  & $4.0^{+0.9}_{-0.9}$  \\ \noalign{{\smallskip}} 
J1730$-$2304 & $2.3^{+3.1}_{-1.8}$  & $-16.0^{+2.7}_{-2.7}$  & $2.5^{+0.7}_{-0.5}$  & $-13.5^{+0.2}_{-0.3}$  & --  & --  & --  & --  & --  & --  & $7.5^{+0.7}_{-0.7}$  \\ \noalign{{\smallskip}} 
J1744$-$1134 & $2.3^{+2.7}_{-1.3}$  & $-15.9^{+2.2}_{-2.8}$  & $3.2^{+0.9}_{-0.6}$  & $-14.2^{+0.2}_{-0.3}$  & $1.4^{+1.8}_{-0.6}$  & $-13.7^{+0.2}_{-3.3}$  & --  & --  & --  & --  & $5.1^{+0.5}_{-0.5}$  \\ \noalign{{\smallskip}} 
J1824$-$2452A & $5.1^{+0.7}_{-0.5}$  & $-13.1^{+0.2}_{-0.2}$  & $2.6^{+0.2}_{-0.2}$  & $-12.43^{+0.04}_{-0.04}$  & --  & --  & --  & --  & --  & --  & $7.9^{+0.7}_{-0.7}$  \\ \noalign{{\smallskip}} 
J1832$-$0836 & $3.2^{+2.5}_{-2.2}$  & $-16.9^{+2.1}_{-2.1}$  & $4.5^{+1.3}_{-1.1}$  & $-13.5^{+0.3}_{-0.4}$  & --  & --  & --  & --  & --  & --  & $2.1^{+2.7}_{-1.5}$  \\ \noalign{{\smallskip}} 
J1857+0943 & $4.9^{+1.4}_{-1.6}$  & $-14.7^{+0.8}_{-0.8}$  & $2.4^{+0.5}_{-0.4}$  & $-13.3^{+0.1}_{-0.1}$  & --  & --  & --  & --  & --  & --  & $8.0^{+4.5}_{-4.2}$  \\ \noalign{{\smallskip}} 
J1902$-$5105 & $3.4^{+2.4}_{-2.3}$  & $-16.1^{+2.6}_{-2.7}$  & $1.4^{+1.6}_{-0.9}$  & $-13.1^{+0.2}_{-0.3}$  & --  & --  & --  & --  & --  & --  & $5.2^{+6.7}_{-3.8}$  \\ \noalign{{\smallskip}} 
J1909$-$3744 & $4.0^{+0.9}_{-0.8}$  & $-14.7^{+0.3}_{-0.8}$  & $2.0^{+0.2}_{-0.1}$  & $-13.66^{+0.04}_{-0.04}$  & $0.6^{+3.5}_{-0.5}$  & $-14.5^{+0.2}_{-1.4}$  & --  & --  & $0.7^{+0.6}_{-0.4}$  & $-13.7^{+0.1}_{-0.3}$  & $4.1^{+0.4}_{-0.4}$  \\ \noalign{{\smallskip}} 
J1933$-$6211 & $3.4^{+2.4}_{-2.3}$  & $-16.3^{+2.4}_{-2.5}$  & $3.3^{+2.4}_{-2.3}$  & $-16.9^{+2.1}_{-2.1}$  & --  & --  & --  & --  & --  & --  & $6.6^{+6.6}_{-4.6}$  \\ \noalign{{\smallskip}} 
J1939+2134 & $6.2^{+0.6}_{-0.7}$  & $-14.6^{+0.3}_{-0.3}$  & $2.8^{+0.2}_{-0.2}$  & $-12.91^{+0.04}_{-0.04}$  & --  & --  & $1.2^{+0.5}_{-0.5}$  & $-13.9^{+0.1}_{-0.1}$  & $2.2^{+2.9}_{-1.5}$  & $-16.3^{+2.6}_{-2.5}$  & $7.3^{+5.9}_{-4.7}$  \\ \noalign{{\smallskip}} 
J2124$-$3358 & $4.7^{+1.5}_{-1.9}$  & $-14.9^{+1.0}_{-0.9}$  & $3.0^{+2.5}_{-2.1}$  & $-17.3^{+2.0}_{-1.8}$  & --  & --  & --  & --  & --  & --  & $6.0^{+2.0}_{-2.0}$  \\ \noalign{{\smallskip}} 
J2129$-$5721 & $3.4^{+2.4}_{-2.3}$  & $-16.6^{+1.9}_{-2.3}$  & $3.1^{+0.5}_{-0.4}$  & $-13.7^{+0.1}_{-0.1}$  & --  & --  & --  & --  & --  & --  & $4.6^{+3.3}_{-2.8}$  \\ \noalign{{\smallskip}} 
J2145$-$0750 & $4.2^{+1.7}_{-1.6}$  & $-14.5^{+0.8}_{-0.9}$  & $1.8^{+0.5}_{-0.3}$  & $-13.5^{+0.1}_{-0.2}$  & --  & --  & --  & --  & --  & --  & $5.8^{+0.7}_{-0.7}$  \\ \noalign{{\smallskip}} 
J2241$-$5236 & $3.0^{+1.7}_{-1.3}$  & $-14.5^{+0.5}_{-3.6}$  & $2.7^{+0.4}_{-0.3}$  & $-14.0^{+0.1}_{-0.1}$  & --  & --  & --  & --  & --  & --  & $4.3^{+0.9}_{-0.9}$  \\ \noalign{{\smallskip}} 

\noalign{{\medskip}} 

\hline\hline
\end{tabular}
}
\label{tab:noise_params}
\end{table*}

\subsection{Achromatic noise processes}

\begin{figure}
\centerline{\includegraphics[width=0.5\textwidth, trim= 0 0 0 0 , clip]{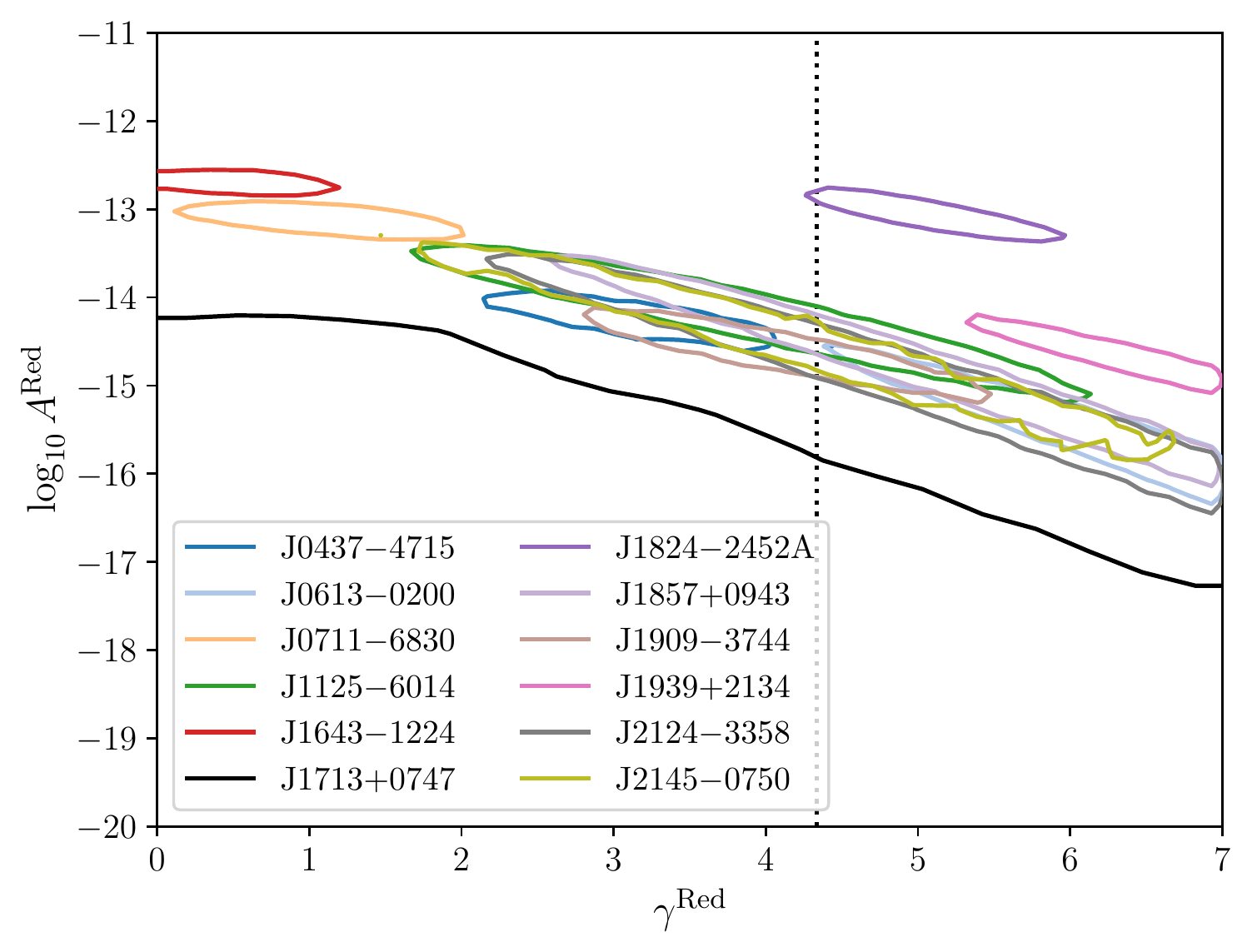}}
\caption{Red-noise posterior probability distributions. Each pulsar with constrained red-noise spectral properties is shown with a colored contour of the maximum likelihood 68\% credibility region. Additionally, the 90\% (upper, one-sided) credible interval for PSR~J1713$-$0747 is shown. The dotted line indicates $\gamma = 13/3$, corresponding to the expected spectral index of GWs produced by an ensemble of SMBHBs.}
\label{fig:red}
\end{figure}

It is critical to understand achromatic processes (those that do not depend on observing frequency), as the GWB is expected to be achromatic. Therefore, GW-induced fluctuations in pulsar timing residuals will be correlated with other achromatic processes (such as spin noise). We include at least one red power-law process to describe the achromatic noise in all pulsars. The posterior probability distribution for this process is constrained at the $>1\sigma$ level for 10 pulsars, shown in Figure \ref{fig:red}. We observe shallow-spectrum noise in PSR~J1643$-$1224 ($\gamma^{\rm Red} = 0.6\pm 0.4$) and PSR~J0711$-$6830 ($\gamma^{\rm Red} = 1.2^{+0.7}_{-0.6}$) and loud, steep-spectrum noise in the relatively 
high magnetic field PSR~J1939+2134 ($\gamma^{\rm Red} = 6.2^{+0.6}_{-0.7}$) and the globular cluster PSR~J1824-2452A ($\gamma^{\rm Red} = 5.1^{+0.7}_{-0.6}$). The noise properties of the remaining pulsars with constrained noise properties are broadly consistent, and the probability-weighted mean of these posteriors is related to the recovered common-spectrum noise in our companion analysis.

\begin{figure}
\centerline{\includegraphics[width=0.5\textwidth, trim= 0 0 0 0 , clip]{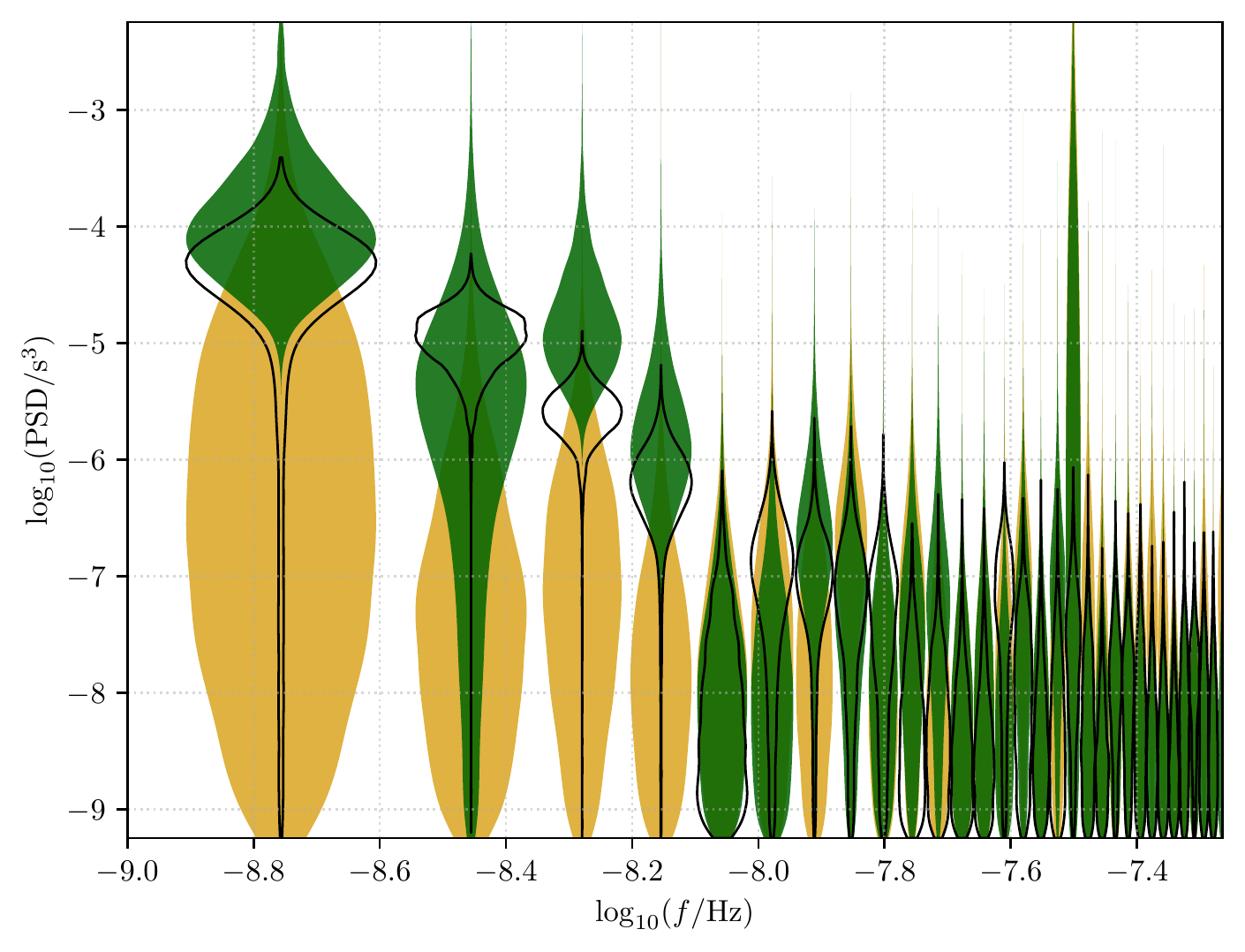}}
\caption{Free-spectrum inference for single-pulsar achromatic red noise, for J1909$-$3744 (green violins) and J1713$+$0747 (gold violins). The black violins show the inferred free spectral model of the common noise detailed in our GW companion paper \citep{PPTA-DR3_gwb}. The violins represent the probability density of the free-spectrum parameter $\rho$, with broader segments of a violin corresponding to higher probability density (with a linear scale). The posteriors are unconstrained at $f=1/(1\,\mathrm{yr}) = 31.7$\,nHz because of degeneracy with the timing model parameters for the pulsar positions.}
\label{fig:freespec}
\end{figure}

All other pulsars show unconstrained spectral properties but are consistent with the population of noise. The exception is PSR~J1713$+$0747, which appears to have both a low amplitude and a shallow spectrum. The 90\% credible interval for the achromatic noise in this pulsar is shown in Figure \ref{fig:red}, and does not intersect with the 1$\sigma$ contours of the other pulsars. This highlights that this pulsar is the most in tension with the presence of a common-spectrum process in the remaining pulsars. To analyze this in further detail, we show a free spectral inference (where the amplitudes of each Fourier frequency are free parameters instead of being constrained to a power law) for the achromatic noise in PSR~J1909$-$3744 and PSR~J1713$+$0747 in Figure \ref{fig:freespec}, along with the free spectral inference of the common-spectrum process \citep{PPTA-DR3_gwb}. This achromatic noise in PSR~J1713$+$0747 is shallow across the lowest frequency bins, explaining the tension with a $\gamma=13/3$ process. This tension with the common noise can also be observed by simply inspecting the timing residuals that are remarkably flat \citep[see Figure 5 in ][]{PPTA-DR3_data}. The weighted rms of the timing residuals after subtracting all frequency-dependent fluctuations (achromatic residuals) is 140\,ns, which is just 22\,ns more than the value of the fully whitened residuals presented in Figure \ref{fig:residuals}. By comparison, PSR~J1909$-$3744 has weighted rms values of 292\,ns and 101\,ns in the achromatic and whitened residuals, respectively.

For the pulsars with HFF noise processes, we observe that the recovered parameters generally have a shallow spectrum, as expected. The spectrum of this process is not constrained to be shallow, but if it is not observed as such, it is completely degenerate with the usual red-noise power law and thus not required for our GW searches.

\subsection{The interstellar medium and solar wind}
\label{sec:iism}

\begin{figure*}
\centerline{\includegraphics[width=\textwidth, trim= 0 0 0 0 , clip]{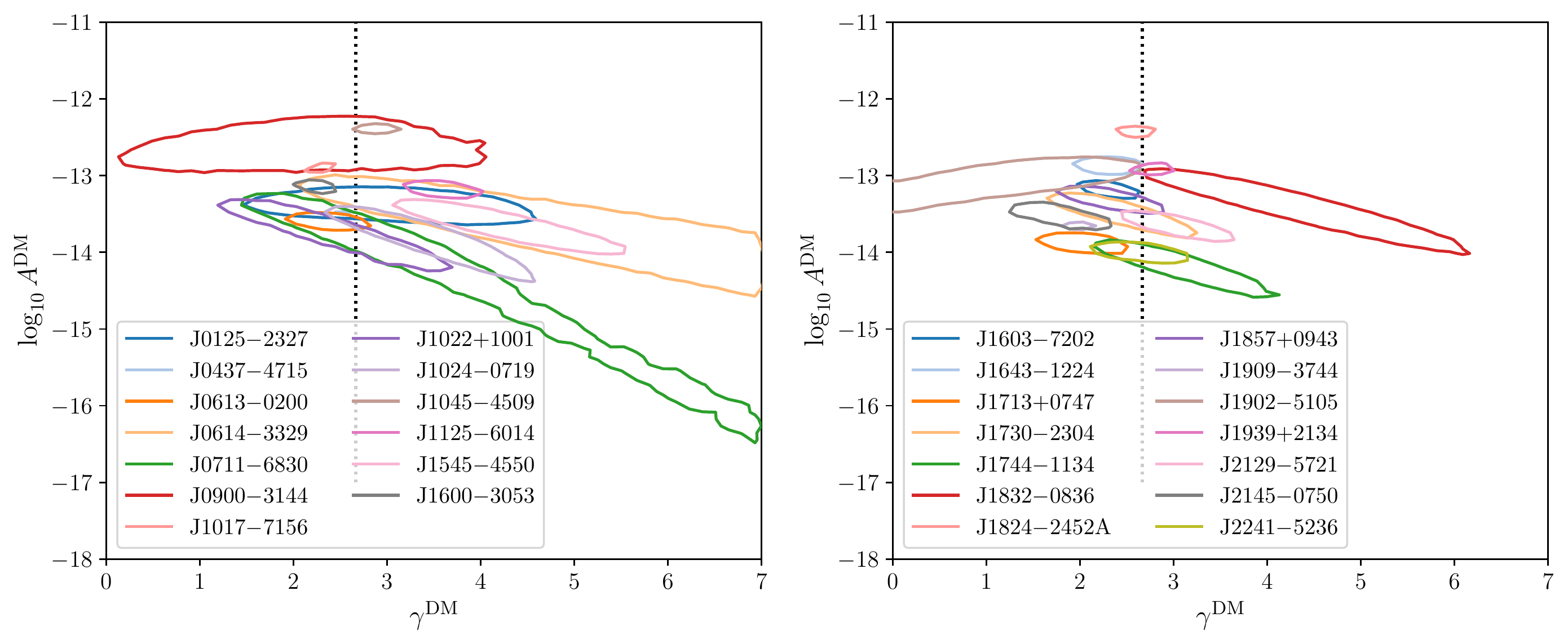}}
\caption{DM Gaussian process posterior probability distributions. Each pulsar with constrained spectral properties for DM variations is shown with a colored contour of the maximum likelihood 68\% credibility region. The dotted line indicates $\gamma = 8/3$, corresponding to a uniform IISM of Kolmogorov turbulence. To aid visual clarity, we divide the pulsars across both panels by right ascension.}
\label{fig:DMvar}
\end{figure*}

The dominant and most widespread effect of the IISM is that of frequency dispersion. Small physical variations in the density of the IISM result in stochastic variations of the dispersive time delay as the pulsar--Earth line of sight changes, with a power spectrum that should depend on the phase structure function of the medium \citep{Foster+90, Rickett90}. The relative motion may also induce periodic (e.g. annual) or nonstationary variations to the DM. We include one power-law Gaussian process model to describe stochastic DM variations for each pulsar. We also include first and second DM time derivative terms in the pulsar timing models, which has the effect of pre-whitening the DM variations \citep{Coles+11, IPTA_dr1_noise}. The maximum likelihood 68\% credible intervals for the amplitude and spectral index of the DM Gaussian process are shown in Figure \ref{fig:DMvar}. The pulsars as a population are broadly consistent with the $\gamma^{\rm DM} = 8/3$ expected from homogeneous Kolmogorov turbulence in the IISM \citep{You+07, Keith+13, IPTA_dr1_noise}. However, individual lines of sight vary about this value, which is expected because of IISM inhomogeneity or anisotropy along individual lines of sight \citep{Rickett90}. For pulsars with large DM and/or precise timing, we also search for the presence of additional time delays resulting from scattering variations that scale as $\nu^{-4}$. We observe significant stochastic variations in the scattering noise for seven pulsars, with the amplitudes and spectral indices summarized in Table \ref{tab:noise_params}.

\begin{figure*}
\centerline{\includegraphics[width=\textwidth, trim= 0 0 0 0 , clip]{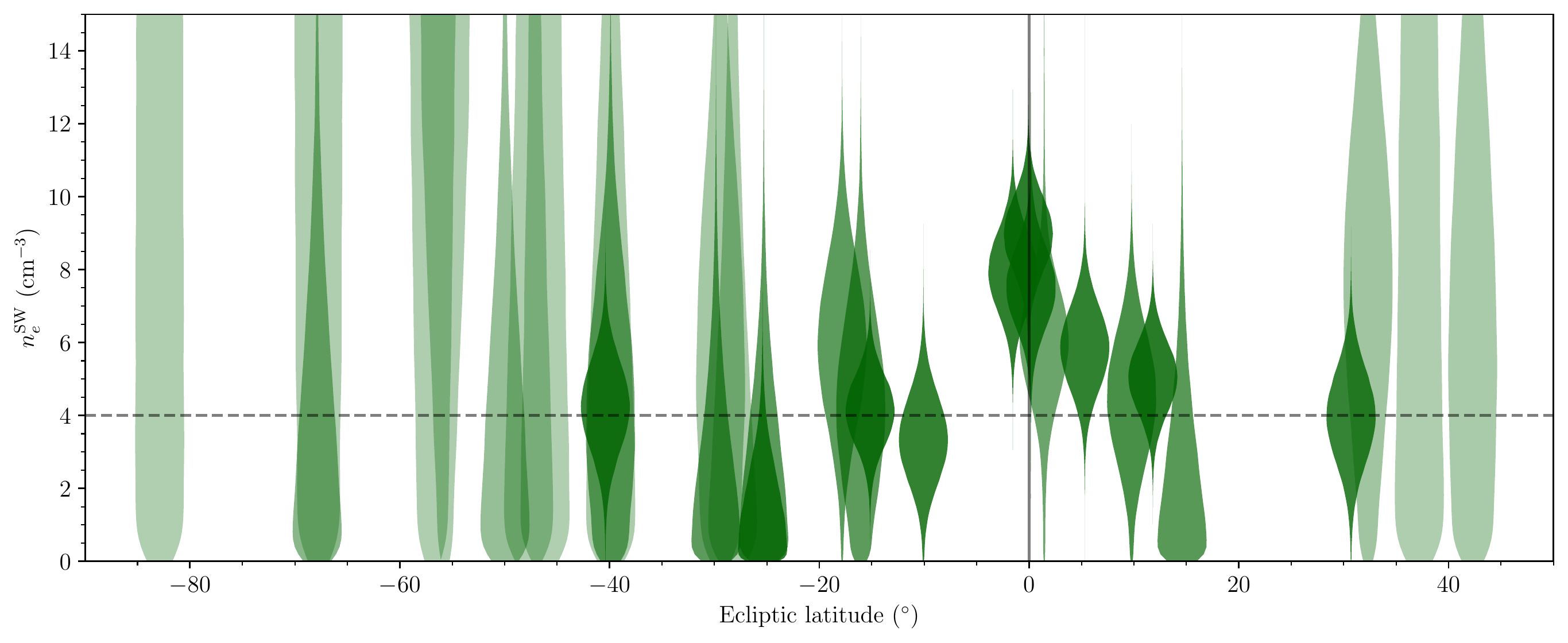}}
\caption{Mean solar wind density at 1\,au ($n^{\mathrm{SW}}_{e}$), as a function of ecliptic latitude for each pulsar. The violins represent the posterior probability density of $n^{\mathrm{SW}}_{e}$, and the violin transparency is set by the width of the distribution of $n^{\mathrm{SW}}_e$ for the corresponding pulsar. The dashed black horizontal line shows the fiducial value of $n^{\mathrm{SW}}_e = 4\,\text{cm}^{-3}$ used when developing the initial timing model \citep{Edwards+06}. The solid vertical line highlights zero ecliptic latitude. While it is a reasonable description for pulsars with $|\text{ELAT}|\gtrsim 15^\circ$, we observe excess electron density for pulsars at low ecliptic latitudes, consistent with annual variations dominated by the slow component of the solar wind \cite[e.g.][]{porowski+22}.}
\label{fig:nearth}
\end{figure*}

For each pulsar we also determined the mean solar wind density, measured at a distance of 1\,au from the Sun, $n^{\mathrm{SW}}_{e}$, and searched for stochastic density variations with time \citep{Hazboun+22}. The posterior probability distributions for the $n^{\mathrm{SW}}_{e}$ from each pulsar are shown in Figure \ref{fig:nearth}. We clearly observe a preference for a higher mean solar wind density for pulsars at low ecliptic latitudes. The solar wind is not spherically symmetric, with fast and slow phases, and with higher densities typically observed at low heliolatitudes \citep[e.g.][]{kojima+98,porowski+22}, which is consistent with what we observe. Pulsars at high ecliptic latitude are insensitive to the mean solar wind density in the search range ($0\,{\rm cm}^{-3} < n^{\mathrm{SW}}_{e} < 20\,{\rm cm}^{-3}$). 

For any pulsars with an unconstrained $n^{\mathrm{SW}}_{e}$ posterior, we use the constant value of $n^{\mathrm{SW}}_{e} = 4\,{\rm cm}^{-3}$ for our GWB noise model \citep[the default value in \textsc{tempo2}][]{Edwards+06}. Ten pulsars have inferred nonzero stochasticity in their solar wind densities: PSR~J0437$-$4715, PSR~J0900$-$3144, PSR~J1022$+$1001, PSR~J1024$-$0719, PSR~J1643$-$1224, PSR~J1713$+$0747, PSR~J1730$-$2304, PSR~J1744$-$1134, PSR~J1909$-$3744, and PSR~J2145$-$0750. However, the spectral properties are poorly constrained with the exception of PSR~J1744$-$1134 and PSR~J1909$-$3744, which have $\gamma^{\rm SW} = 1.6^{+0.5}_{-0.7}$ and $\gamma^{\rm SW} = 0.9^{+0.5}_{-0.4}$ respectively (median and 68\% credible interval). This low-frequency power that we observe may be associated with density variations due to the solar cycle. We encourage further development in this framework to fully capture the complexity of solar wind density variations \citep[e.g.,][]{You+07,Tiburzi+21}, and to harness MSPs as tools for studying the heliosphere \citep{Madison+19,Kumar+22}.

\subsection{Timing offsets}

\begin{figure}
\centerline{\includegraphics[width=0.5\textwidth, trim= 0 0 0 0 , clip]{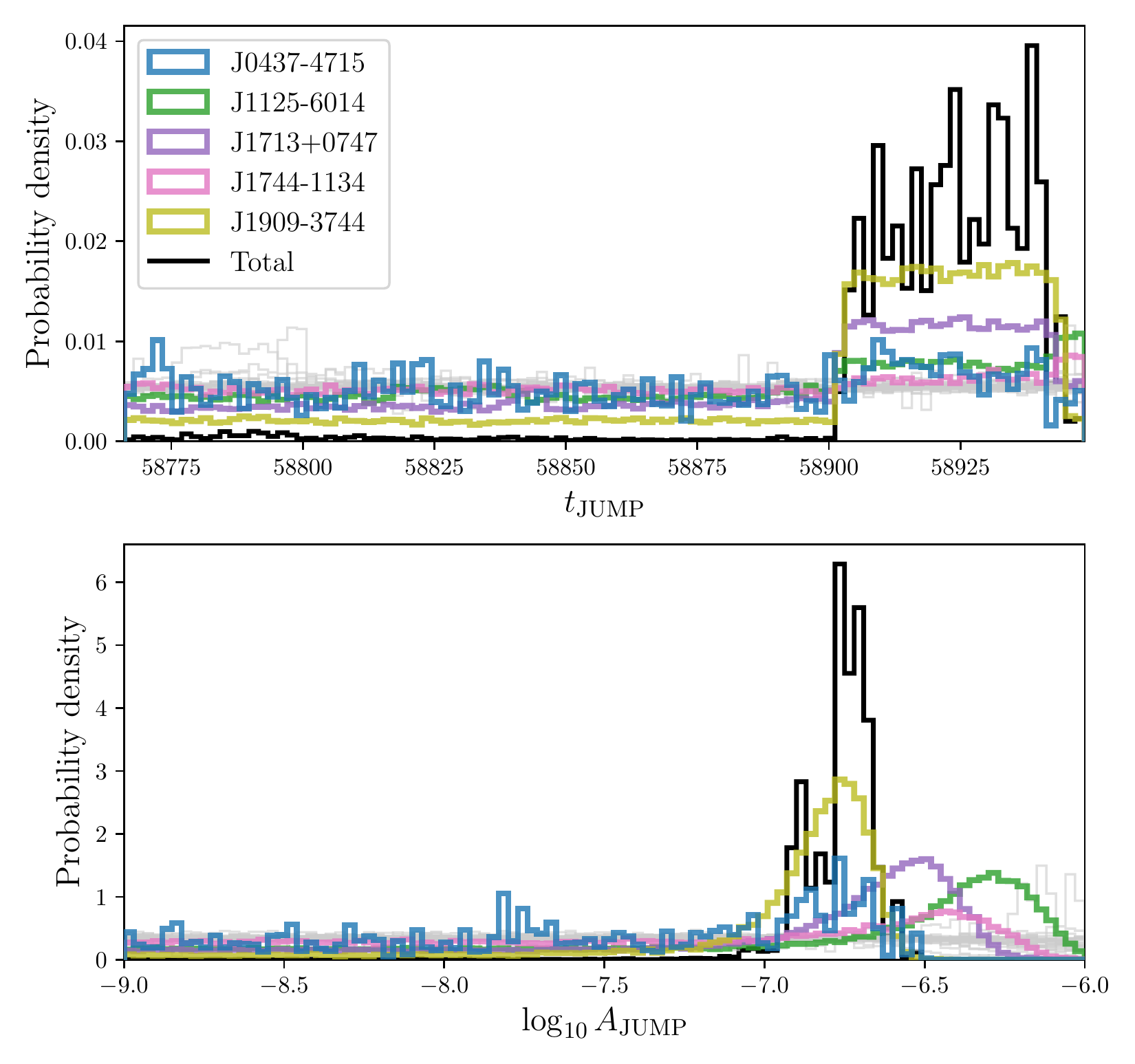}}
\caption{Factorized likelihood analysis for timing offset parameters $t_\mathrm{JUMP}$ (MJD; top) and $\log_{10} A_\mathrm{JUMP}$ (with $A_\mathrm{JUMP}$ in seconds; bottom). We color the marginal posteriors for pulsars with positive support for the MJD 58925 timing offset and show the marginal posteriors for other pulsars in light gray. In the bottom panel, we have filtered $\log_{10} A_\mathrm{JUMP}$ samples that coincide with the time span with positive support for the MJD 58925 timing offset.}
\label{fig:jump}
\end{figure}

We identify one significant timing offset in the Medusa UWL data that is supported by multiple pulsars. The epoch for this system offset is approximately MJD~58925, and its magnitude is $\mathcal{O}(100\,{\rm ns})$. The factorized likelihood analysis for the epoch and amplitude of this offset is shown in Figure \ref{fig:jump}. We found that this epoch coincides with a system digitizer resynchronization, which may be its cause. Unfortunately, it is not currently possible to predict the magnitude of the offset except with the data themselves, and we therefore allow it to vary in the presence of our noise models. The epoch is fixed for our analyses, but the amplitude is considered a nuisance parameter and is analytically marginalized as part of the pulsar timing model.

\subsection{Band and system noise}
\label{sec:bandgroup}

Significant band noise at $\nu \leq 960\,$MHz was only observed in PSR~J1643$-$1224 ($\log\mathcal{B} = 9.7$), PSR~J1713$+$0747 ($\log\mathcal{B} = 6.2$), and PSR~J1909$-$3744 ($\log\mathcal{B} = 1.6$). For PSR~J1643$-$1224, the source of band noise may be its known scattering variations \citep[][see also Section \ref{sec:iism}]{Ding+23}, which may not be well described by a power law \citep{Mall+22}.

\begin{figure}
    \centering
    \includegraphics[width=\linewidth]{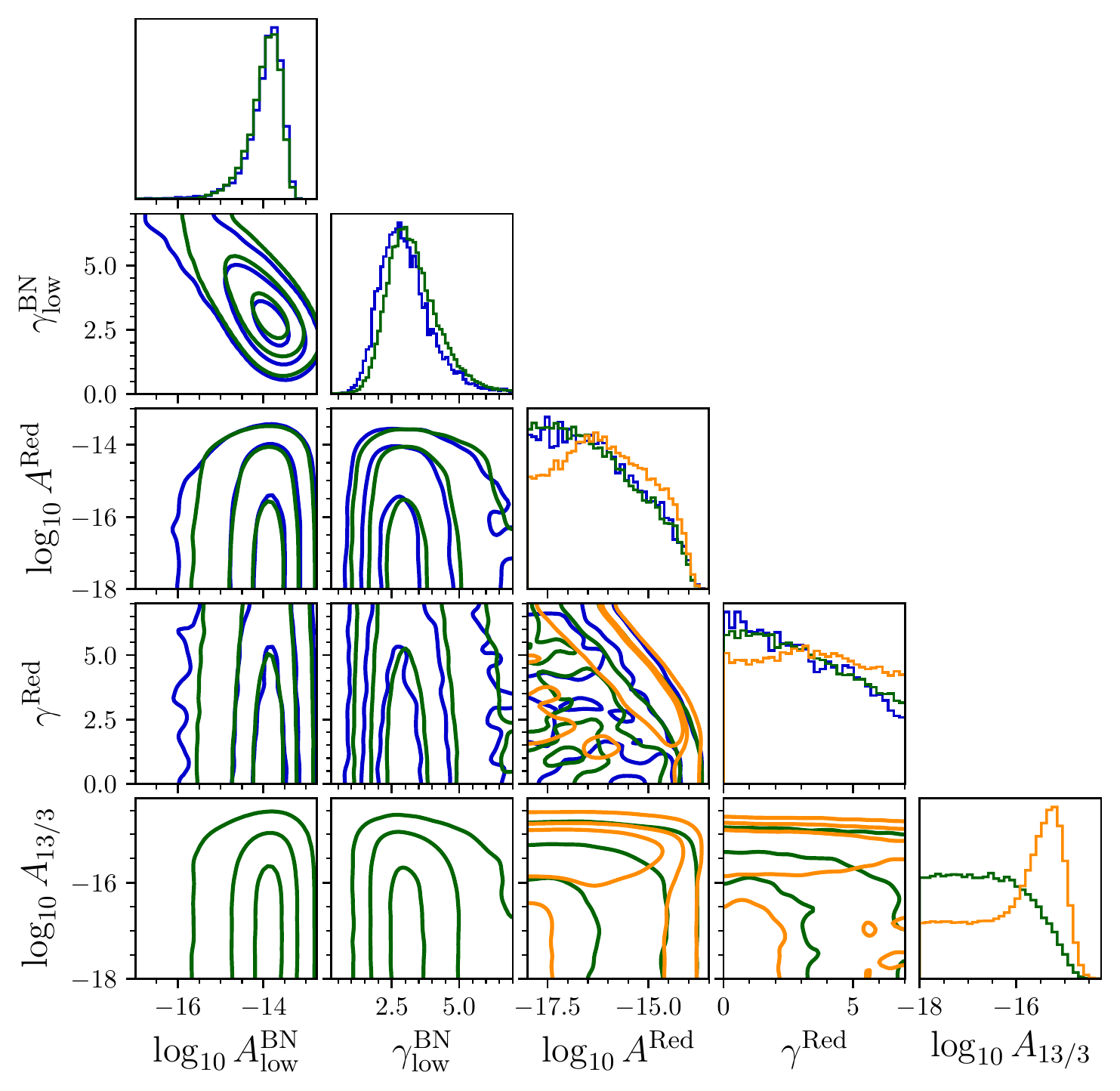}
    \caption{Leakage of low-frequency ($\nu \leq 960\,$MHz) band noise into achromatic red noise and $\gamma=13/3$ red noise in PSR~J1713$+$0747. Green contours show posteriors from our single-pulsar model runs with a $\gamma=13/3$ process. Blue contours show the posteriors assuming a fixed-amplitude $\gamma=13/3$ process corresponding to the recovered common-spectrum noise, from a pairwise correlation analysis \citep{PPTA-DR3_gwb}. Orange contours show posteriors with a single-pulsar noise model including a 13/3 process, without a low-frequency band-noise term in the model.}
    \label{fig:1713bandnoise}
\end{figure}

We find that after including band noise in the model for PSR~J1713$+$0747, the achromatic red-noise properties change. The amplitude of noise assuming a spectral index of $\gamma=13/3$ is also larger without band noise, but we attribute this to mis-specification rather than truly achromatic noise and/or common noise. Indeed, when we assume an additional red-noise process at a fixed amplitude and spectral index, corresponding to the inferred common-spectrum noise, we infer a near-identical posterior distribution for the low-frequency band noise, as shown in Figure \ref{fig:1713bandnoise}. 

We find evidence for excess noise in some of the early signal processing systems used by the PPTA. This includes strong evidence for excess noise in the \textsc{WBCORR} system in the 10\,cm band for PSR~J1713$+$0747, with $\log\mathcal{B} = 4.5$. Excess noise was also identified in this system with the PPTA-DR2 analysis of PSR~J0437$-$4715 \citep{PPTA_dr2_noise}, but these data were not included as part of the PPTA-DR3. The Savage--Dickey Bayes factors for system noise in \textsc{CASPSR} 40\,cm ($\log_{10} A = -14.39 \pm 0.6$ and $\gamma = 4.8^{+1.4}_{-1.6}$) and UWL \textsc{PDFB4} 20\,cm ($\log_{10} A = -14.1^{+0.3}_{-0.5}$ and $\gamma = 4.0^{+2.0}_{-1.4}$) data for PSR~J0437$-$4715 can only be estimated as $\log\mathcal{B} > 12$ because of a limited number of samples in the low-amplitude tail of their marginal posteriors. We also found significant system noise in \textsc{UWL\_sbA} for PSR~J1017$-$7156 ($\log_{10} A = -13.5\pm0.5$ and $\gamma = 3.9^{+2.1}_{-1.9}$, with $\log \mathcal{B} =5.4$.

System noise terms applying to various other UWL subbands were included in our model for some pulsars because they have partially constrained posterior probability distributions (though low significance, $\log\mathcal{B}\sim 1$) in this data set. These may indicate subthreshold systematics, which, while small in individual pulsars, may result in nonnegligible amounts of excess noise that potentially could be erroneously attributed to other terms in a joint noise analysis. These additional system noise terms, per pulsar, were 
{UWL\_sbA} and {UWL\_sbG} (PSR~J0437$-$4715);
{UWL\_sbA} and {UWL\_sbD} (PSR~J1017$-$7156);
{UWL\_sbE} and {UWL\_sbH} (PSR~J1022$+$1001);
{UWL\_sbA}, {UWL\_sbE}, and {UWL\_sbF} (PSR~J1713$+$0747).

\subsection{Noise and timing model validation}

\begin{figure*}
\includegraphics[width=\textwidth]{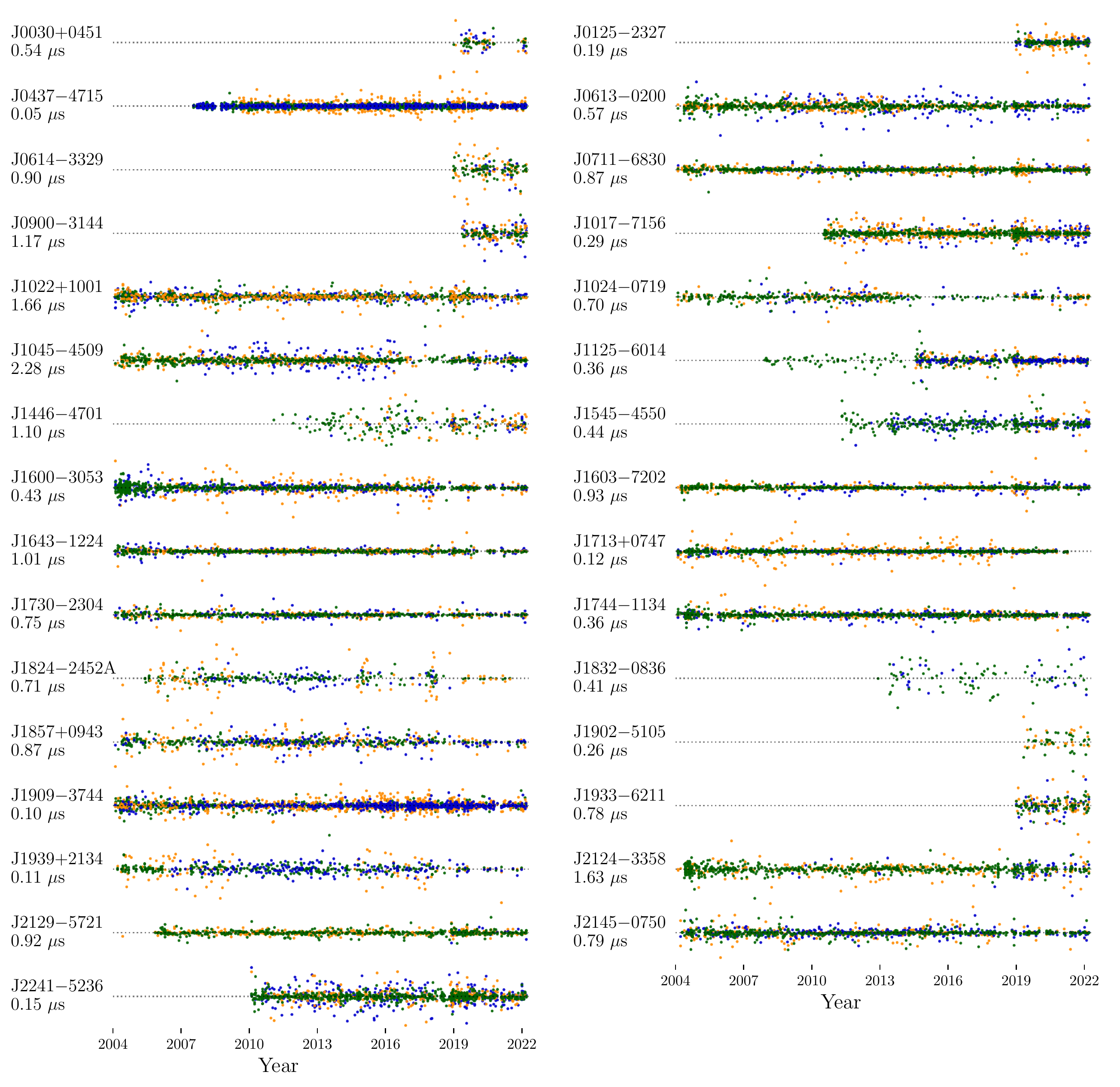}
\caption{Whitened, band-averaged timing residuals for the PPTA-DR3. The 10, 20, and 40\,cm wave bands are shown in blue, green, and orange, respectively. We show the weighted rms of the whitened residuals in the 20\,cm band beneath each pulsar label.}
\label{fig:residuals}
\end{figure*}

The noise models were checked for completeness by analyzing the whitened and normalized residuals in each frequency band using a Lomb--Scargle periodogram and an Anderson--Darling test. The whitened (but not normalized) residuals and their band-averaged weighted rms values are shown in Figure \ref{fig:residuals}. We found that $2.3\%$ of the periodogram (fluctuation) frequencies of interest ($f_t < 1/240\,{\rm days}$) contain excess power at the 2$\sigma$ level and $0.47\%$ contain excess power at the 3$\sigma$ level. Only PSR~J1022$+$1001 showed excess power at the $>3 \sigma$ level in multiple observing bands, at the frequency nearest to $f_t = 1/(1\,{\rm yr})$. Achromatic noise at this frequency is subtracted when fitting for position in the timing model, so the excess noise must be chromatic (and therefore is unlikely to impact our achromatic common-spectrum noise search). It could be related to the solar wind, as this pulsar has the lowest ecliptic latitude ($-0.063\,$degrees) of the pulsars in our data set. Alternatively, the excess chromatic noise may be related to profile variations, but the coincidence with $f_t = 1/(1\,{\rm yr})$ suggests a solar system origin. No periodogram frequencies were outliers at the 4$\sigma$ level in any pulsar, and overall the whitened residuals are consistent with white noise.

Next, we tested for Gaussianity of the whitened and normalized residuals in each frequency band (and band-averaged) using the ADS. After our noise modeling was complete, all pulsars passed the ADS test for Gaussianity. During the course of our analysis, PSR~J0125$-$2327, PSR~J0614$-$3329, PSR~J1902$-$5105, and PSR~J2241$-$5236 initially failed the ADS test, which motivated an inspection of their timing and noise models. We identified Shapiro delays in PSR~J0614$-$3329 and PSR~J1902$-$5105, and a secular advance of the projected semi-major axis ($\dot{x}$) in PSR~J0125$-$2327. We also found that PSR~J2241$-$5236 required 17 orbital frequency derivative parameters to accurately describe the instability of its orbit with a nondegenerate companion \citep[an increase over the 10 used for the previous PPTA noise and timing analyses; ][]{PPTA_dr2_noise, Reardon+21}. 

From the Shapiro delay of PSR~J0614$-$3329, we find that the orbit is extremely edge-on with $\sin i = 0.99965 \pm 0.00046$, where $i$ is the orbital inclination angle. The companion mass is measured as $M_c = 0.26 \pm 0.04\, M_\odot$, and the pulsar mass is derived from the binary mass function to be $M_p = 1.2 \pm 0.3\, M_\odot$. The Shapiro delay for PSR~J1902$-$5105 lacks the precision necessary to derive meaningful pulsar and companion masses, and we defer a detailed analysis of its timing model to future work. The $\dot{x}$ measurement for PSR~J0125$-$2327 gives a constraint on the system inclination angle, $i \leq 37^\circ$ with 95\% confidence \citep[for details see, e.g., ][]{Sandhu+97, Reardon+21}.

After updating the timing models and reevaluating the noise parameters, each of these pulsars passed the ADS tests, indicating that this statistic is sensitive to some timing model errors. 

\section{Discussion} 
\label{sec:discussion}

The noise processes observed in MSP TOAs have been shown to be more complex than previously assumed. Rather than the low-frequency fluctuations resulting only from irregularities in the pulsar spin or dispersion by the interstellar medium, multiple PTA pulsars have also demonstrated timing irregularities due to sudden changes in their magnetospheres and interstellar scattering. Errors in the models for the various processes and irregularities in observing systems induce further noise processes, occasionally in discrete observing bands. We have introduced noise terms into our pulsars based on published inferences from earlier data sets and predictions for which processes will be present.

While characterizing the stochastic and deterministic signals in pulsar timing residuals is astrophysically interesting in its own right, our main motivation for constructing detailed noise models is ultimately to minimize the possibility of misattributing signals to a common-spectrum stochastic process. Failure to include low-frequency band noise in the model for PSR~J1713$+$0437 induces some (weak) support for steep-spectrum achromatic noise (Figure \ref{fig:1713bandnoise}), which would contribute positively to a common-noise search. However, including the common noise in the model for PSR~J1713+0747 (with fixed amplitude and spectral index) does not change the inference of low-frequency band noise, indicating that the two processes are distinct. Instead, we observe that including this common noise induces a quadratic-like structure into the timing residuals, and as a result the inferred spin frequency derivative changes from $\dot{f} = -4.083914(7)\times 10^{-16}\,{\rm s}^{-2}$ (assuming our maximum likelihood red noise for J1713$+$0747) to $\dot{f} = -4.08381(3)\times 10^{-16}\,{\rm s}^{-2}$ (assuming the inferred common-spectrum noise). A careful analysis of the interaction of timing model parameters with the various noise processes would be valuable for future analyses.

To demonstrate the efficacy of our detailed noise modeling for searches for the stochastic GWB, we compared our full noise model with a simplified version that still captures the bulk of the observed noise, albeit with lower accuracy. We performed parameter estimation for an uncorrelated common-spectrum stochastic process using basic single-pulsar noise models, ignoring the higher-order stochastic single-pulsar terms. The basic noise models contained only the Gaussian process models for achromatic red noise and DM variations and deterministic terms (magnetospheric events, DM events, and a spherically symmetric $n^{\mathrm{SW}}_e = 4\,\text{cm}^{-3}$ solar wind model). Relative to this model, our advanced noise modeling included additional terms for band, system, HFF, and chromatic noise, along with variable solar wind terms. 

\begin{figure}
    \centering
    \includegraphics[width=\linewidth]{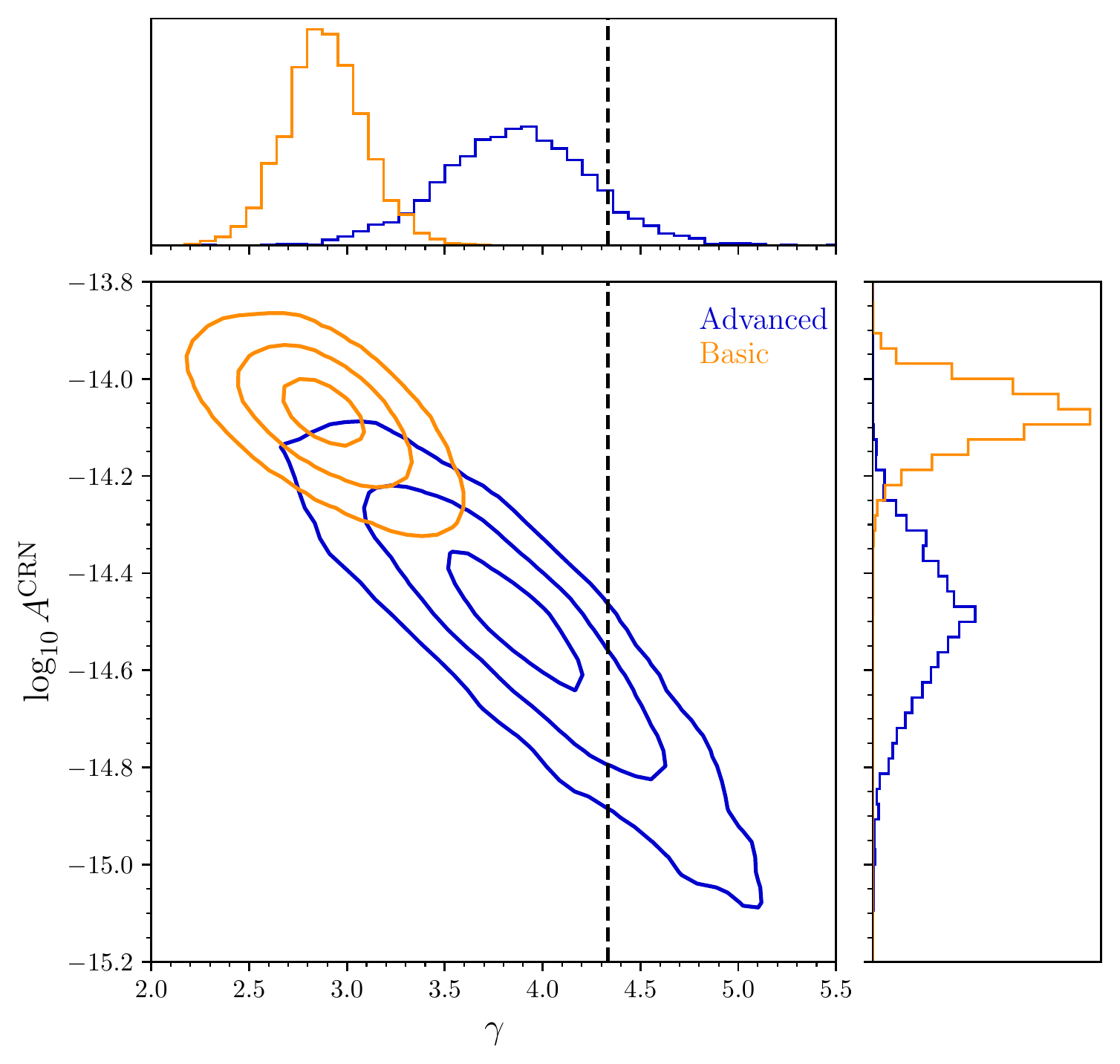}
    \caption{Marginal posterior probability distributions for the measured logarithmic amplitude and spectral index of a common uncorrelated process under basic (orange) and advanced (blue) noise models. The contours on the two-dimensional distribution show the 1$\sigma$, 2$\sigma$, and 3$\sigma$ credible intervals for each model.}
    \label{fig:crn_comp}
\end{figure}

A comparison of the recovered properties of the common-spectrum noise is shown in Figure \ref{fig:crn_comp}. We find that the choice of single-pulsar noise models significantly influences the recovered spectral characteristics of a common process -- the basic noise model returns $\log_{10}A^{\rm CRN} = -14.08 \pm 0.06$ and $\gamma = 2.9 \pm 0.2$, while the advanced noise model returns $\log_{10} A^{\mathrm{CRN}} = -14.50^{+0.14}_{-0.16}$ and $\gamma = 3.87 \pm 0.36$ \citep{PPTA-DR3_gwb}.

\begin{figure}
    \centering
    \includegraphics[width=\linewidth]{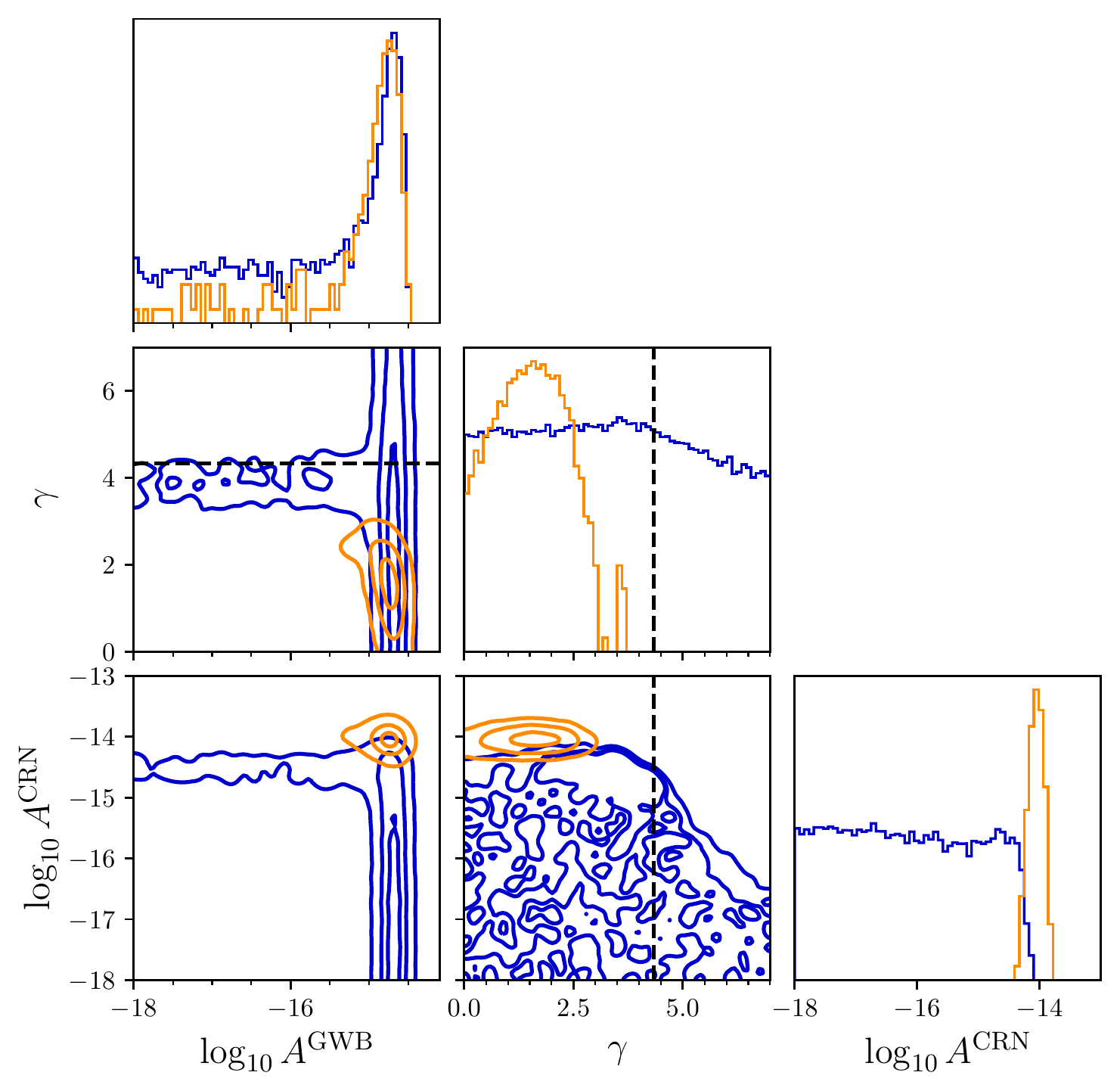}
    \caption{Marginal posterior probability distributions for the measured logarithmic amplitude ($\log_{10} A^{\mathrm{CRN}}$) and spectral index ($\gamma$) of a common uncorrelated process, along with the logarithmic amplitude of an HD-correlated fixed spectral index ($\gamma=13/3$) common process, under basic (orange) and advanced (blue) noise models. The single-parameter marginal distributions are shown with logarithmic $y$-axes, to highlight the low-density tails of the distributions.}
    \label{fig:crn_hd_comp}
\end{figure}

We also performed parameter estimation for a two-component common-spectrum process model: one as an HD spatially correlated stochastic process with spectral index fixed at $\gamma =13/3$, along with a common uncorrelated stochastic process with a free spectral index. We analyzed these models under both the basic and detailed single-pulsar noise models. Our motivation was to investigate whether unmodeled noise in pulsar timing residuals could be absorbed into the common-spectrum process and whether this influences the characteristics of a prescriptive GWB model (HD spatial correlations and $\gamma=13/3$, although the inclusion of the spatial correlations here does not significantly affect the inference). The results are shown in Figure \ref{fig:crn_hd_comp}. In both cases, we recover an HD-correlated fixed spectral index process amplitude of $\log_{10} A^{\mathrm{GWB}} \approx -14.7$ \citep{PPTA-DR3_gwb}. When using the detailed single-pulsar noise models, we do not detect any additional common process -- both $\gamma$ and $\log_{10} A^\mathrm{CRN}$ are unconstrained. In contrast, when using basic noise models, we recover a significant detection of an additional common-noise component, with $\log_{10} A^\mathrm{CRN} = -14.04^{+0.08}_{-0.07}$ and $\gamma = 1.7^{+0.6}_{-0.7}$. We attribute the presence of this component to mis-specified single-pulsar noise.

We draw three conclusions from this analysis in relation to searches for a common-spectrum process:
\begin{enumerate}
    \item The choice of single-pulsar noise models influences the spectral characteristics of a recovered common-spectrum process. When we employ single-pulsar noise models constructed with the aim of whitening the timing residuals, the spectral characteristics of the common-spectrum process are more consistent with $\gamma=13/3$ expected for a GWB from SMBHBs.
    \item Unmodeled noise in individual pulsars can leak into the common-spectrum process. This can lead to spurious statistical support for a common-spectrum process, similar to the conclusions of \citet{Goncharov+22} and \citet{Zic+22}, and can also bias the recovered spectral characteristics as described above.
    \item Under our detailed single-pulsar noise models, an HD-correlated common noise is a consistent description of the common-spectrum process discussed in \citet{PPTA-DR3_gwb}. We detect no additional uncorrelated common-spectrum process when we include an HD-correlated fixed spectral index common process.
\end{enumerate}

\section{Conclusions}
\label{sec:conclusions}

We have described the construction of single-pulsar noise models for 31 MSPs from the PPTA-DR3. As these noise models form the null hypothesis from which common-spectrum stochastic processes are inferred, a liberal approach was taken regarding the addition of noise terms because we are interested in achieving the most robust common-noise inference possible.

Our noise models include parameters describing white noise, achromatic red noise, and DM variations (originating from the IISM and solar wind) for each pulsar. We also include higher-order terms that describe excess noise present in subsets of the measurements, such as individual frequency bands or observing systems for some pulsars. We have also searched for and corrected unmodeled instrumental timing offsets within our noise modeling process.

We have demonstrated that the advanced noise modeling has a significant impact on the recovered spectral properties in a common-noise (e.g. GWB) search. The common noise is well modeled with a single power law, with a steep spectral index that is consistent with $\gamma=13/3$. With more basic noise models (accounting for achromatic red noise, DM variations, and deterministic events), a similar amplitude is recovered for a process at $\gamma=13/3$, but excess noise described by a shallow spectrum ($\gamma = 1.7^{+0.6}_{-0.7}$) is also present. We suggest that the detection of this distinct uncorrelated common process in addition to a fixed spectral index common process is indicative of incomplete noise modeling. 

Unmodeled noise processes can cause errors in the inference of the achromatic stochastic processes of interest, even when the unmodeled processes are chromatic or band limited. The inclusion of the common-spectrum stochastic process in the model of PSR~J1713$+$0747 does not change the inference of low-frequency band noise. Instead, the spin period derivative and its uncertainty change, suggesting that if the stochastic process truly is common, it is present primarily at frequencies lower than the fundamental frequency for our data set of this pulsar. Longer pulsar timing data sets, for example, from the IPTA, will be able to determine whether this is the case or not. We have also detected new significant timing model parameters in four pulsars, which were required to accurately characterize the observed TOAs. A more detailed timing model analysis for all pulsars is deferred to future work, but noise modeling is clearly an attractive way to detect overlooked timing parameters.

For the common-noise search, the white-noise parameters are fixed at their maximum likelihood values from our analysis, while all other parameters with some level of time correlation ($\sim 260$ of them) are sampled simultaneously with the models of common processes. We are confident that this myriad of parameters lay the groundwork for robust GW inference.

\begin{acknowledgments}

Murriyang, the Parkes 64\,m radio telescope, is part of the Australia Telescope National Facility (\url{https://ror.org/05qajvd42}), which is funded by the Australian Government for operation as a National Facility managed by CSIRO.  We acknowledge the Wiradjuri People as the Traditional Owners of the Observatory site. We acknowledge the Wurundjeri People of the Kulin Nation and the Wallumedegal People of the Darug Nation as the Traditional Owners of the land where this work was primarily carried out. We thank the Parkes Observatory staff for their support of the project for nearly two decades. We also thank CSIRO Information Management and Technology High Performance Computing group for access and support with the petrichor cluster. We acknowledge the use of the Python packages \textsc{numpy} \citep{numpy}, \textsc{scipy} \citep{scipy}, \textsc{chainconsumer} \citep{chainconsumer}, and \textsc{corner} \citep{corner} for parts of this work. Part of this research was undertaken as part of the Australian Research Council (ARC) Centre of Excellence for Gravitational Wave Discovery (OzGrav) under grant CE170100004. R.M.S. acknowledges support through ARC Future Fellowship FT190100155. Work at NRL is supported by NASA. S.D. is the recipient of an Australian Research Council Discovery Early Career Award (DE210101738) funded by the Australian Government. H.M. acknowledges the support of the UK Space Agency, grant No. ST/V002813/1.

\end{acknowledgments}


\end{document}